\documentclass[aps,pra,reprint,amsmath,amssymb,showpacs,nofootinbib,floatfix]{revtex4-1}

\usepackage{graphicx}        
\usepackage{bbm}
\usepackage{amsmath,amssymb,units}
\usepackage[usenames, dvipsnames]{color}
\usepackage[colorlinks,citecolor=Blue,urlcolor=Blue,linkcolor=Blue,hyperindex,breaklinks]{hyperref}

\usepackage{url}

\newcommand{\bra}[1]{\left\langle{#1}\right\vert}
\newcommand{\ket}[1]{\left\vert{#1}\right\rangle}
\newcommand{\braket}[2]{\ensuremath{{\langle #1}|{#2 \rangle}}}
\newcommand{\ketbra}[2]{\ensuremath{|{#1 \rangle}{\langle #2}|}}
\providecommand{\abs}[1]{\left\lvert#1\right\rvert}
\newcommand{\D}{\text{d}}
\newcommand{\E}{\text{e}}
\newcommand{\I}{\text{i}}
\newcommand{\bvec}[1]{\ensuremath{\mathbf{#1}}}
\newcommand{\buvec}[1]{\ensuremath{\mathbf{\hat{#1}}}}

\begin{document}

\title{The quantum-to-classical transition and decoherence}
\author{Maximilian Schlosshauer} 
\affiliation{Department of Physics, University of Portland, 5000 North Willamette Boulevard, Portland, Oregon 97203, USA}

\begin{abstract}
I give a pedagogical overview of decoherence and its role in providing a dynamical account of the quantum-to-classical transition. The formalism and concepts of decoherence theory are reviewed, followed by a survey of master equations and decoherence models. I also discuss methods for mitigating decoherence in quantum information processing and describe selected experimental investigations of decoherence processes.\\[-.3cm]

{\bf Note:} Please see \href{https://arxiv.org/abs/1911.06282}{\texttt{arXiv:1911.06282 [quant-ph]}} (published as \emph{Phys.\ Rep.\ }{\bf 831}, 1--57, 2019) for a much more extensive and up-to-date review of decoherence.
\end{abstract}

\maketitle

\tableofcontents

\section{Introduction}

Realistic quantum systems are never completely isolated from their environment. When a quantum system interacts with its environment, it will in general become entangled with a large number of environmental degrees of freedom. This entanglement influences what we can locally observe upon measuring the system. In particular, quantum interference effects with respect to certain physical quantities (most notably, ``classical'' quantities such as position) become effectively suppressed, making them prohibitively difficult to observe in most cases of practical interest. This is the process of \emph{decoherence}, sometimes also called \emph{dynamical decoherence} or \emph{environment-induced decoherence} \cite{Zeh:1970:yt,Zurek:1981:dd,Zurek:1982:tv,Paz:2001:aa,Zurek:2002:ii,Schlosshauer:2003:tv,Bacciagaluppi:2003:yz,Joos:2003:jh,Schlosshauer:2007:un,Schlosshauer:2019:qd}. Stated in general and interpretation-neutral terms, decoherence describes how entangling interactions with the environment influence the statistics of results of future measurements on the system. 

Formally, decoherence can be viewed as a dynamical filter on the space of quantum states, singling out those states that, for a given system, can be stably prepared and maintained, while effectively excluding most other states, in particular, nonclassical superposition states of the kind popularized by Schr\"odinger's cat. In this way, decoherence lies at the heart of the \emph{quantum-to-classical transition}. It ensures consistency between quantum and classical predictions for systems observed to behave classically. It provides a quantitative, dynamical account of the boundary between quantum and classical physics. In any concrete experimental situation, decoherence theory specifies the physical requirements, both qualitative and quantitative, for pushing the quantum--classical boundary toward the quantum realm. Decoherence is a pure quantum effect, to be distinguished from classical dissipation and stochastic fluctuations (noise).

Decoherence processes are extremely efficient. Even when the environment does not, from a classical point of view, impart significant classical perturbations on the system, quantum-mechanically the system will in most circumstances become rapidly and strongly entangled with the environment. Furthermore, due to the many uncontrollable degrees of freedom of the environment, such entanglement is usually irreversible for all practical purposes. Increasingly realistic models of decoherence processes have been developed, progressing from toy models to complex models tailored to specific experiments (see Sec.~\ref{sec:decmodels}). Advances in experimental techniques have made it possible to observe the gradual action of decoherence in experiments such as matter-wave interferometry \cite{Hornberger:2012:ii}, cavity QED \cite{Raimond:2001:aa}, and superconducting systems \cite{Leggett:2002:uy} (see Sec.~\ref{sec:exper-observ-decoh}). 

The superposition states necessary for quantum information processing are typically also those most susceptible to decoherence. Thus, decoherence is a major barrier to implementing devices for quantum information processing such as quantum computers (see Sec.~\ref{sec:decoh-errcorr}). Qubit systems must be engineered to minimize environmental interactions detrimental to the preparation and longevity of the desired superposition states. At the same time, they must remain sufficiently open to allow for their control. Quantum error correction can undo some of the decoherence-induced degradation of the superposition state and will be an integral part of quantum computers (see Sec.~\ref{sec:corr-decoh-induc}). Not only is decoherence relevant to quantum information, but also vice versa. An information-centric view of quantum mechanics proves helpful in conveying the essence of the decoherence process and is also used in recent explorations of the role of the environment as an information channel (see Sec.~\ref{sec:envir-monit-inform}).

It is a curious ``historical accident'' (Joos's term \cite[p.~13]{Joos:1999:po}) that the role of the environment in quantum mechanics was appreciated only relatively late. While one can find---for example, in Heisenberg's writings \cite{Schlosshauer:2011:un}---a few early anticipatory remarks about the role of environmental interactions in the quantum-mechanical description of systems, it wasn't until the 1970s that the ubiquity and implications of environmental entanglement were realized by Zeh \cite{Zeh:1970:yt,Kubler:1973:ux}. It took another decade for the formalism of decoherence to be developed, chiefly by Zurek \cite{Zurek:1981:dd,Zurek:1982:tv}, and for concrete models and numerical estimates of decoherence rates to be worked out \cite{Joos:1985:iu,Zurek:1986:uz}. 

Review papers on decoherence include Refs.~\cite{Zurek:2002:ii,Paz:2001:aa,Hornberger:2009:aq,Schlosshauer:2003:tv,Schlosshauer:2019:qd}. There are two books on decoherence: a volume by Joos et al.\ \cite{Joos:2003:jh} (a collection of chapters written by different authors) and a monograph by this author \cite{Schlosshauer:2007:un}. Ref.\ \cite{Breuer:2002:oq} also contains material on decoherence. Foundational implications of decoherence are discussed in Refs.~\cite{Bacciagaluppi:2003:yz,Schlosshauer:2003:tv,Schlosshauer:2006:rw,Schlosshauer:2007:un}.

\section{\label{sec:form-basic-conc}Basic formalism and concepts}

In the double-slit experiment, we cannot observe an interference pattern if we also measure which slit the particle went through (that is, if we obtain perfect \emph{which-path information}). In fact, there is a continuous tradeoff between interference (phase information) and which-path information: the better we can distinguish the two possible paths, the less visible the interference pattern becomes \cite{Wooters:1979:az}. What is more, for a decrease in interference visibility to occur it suffices that there are degrees of freedom \emph{somewhere in the world} that, \emph{if they were measured}, would allow us to make, with a certain degree of confidence, a statement about the path of the particle through the slits. While we cannot say that prior to their measurement, these degrees of freedom have encoded information about a particular, definitive path of the particle---instead, we have merely \emph{correlations} involving both possible paths---no actual measurement is required to bring about the decrease in interference visibility. It is enough that, \emph{in principle}, we could make such a measurement to obtain which-path information. 

This is somewhat loose talk, and conceptual caveats lurk. But it captures quite well the essence of what is happening in decoherence, where those ``degrees of freedom somewhere in the world'' are the degrees of freedom of the system's environment interacting with the system, leading to the creation of quantum correlations (entanglement) between system and environment. Decoherence can thus be thought of as a process arising from the \emph{continuous monitoring of the system by the environment}; effectively, the environment is performing nondemolition measurements on the system (see Sec.~\ref{sec:envir-monit-inform}). We now give a formal quantum-mechanical account of what we have just tried to convey in words, and then flesh out the consequences and details.

\subsection{Decoherence and interference damping}

Consider again the double-slit experiment and denote the quantum states of the particle (call it $S$, for ``system'') corresponding to passage through slit 1 and 2 by $\ket{s_1}$ and $\ket{s_2}$, respectively. Suppose that the particle interacts with another system $E$---for example, a detector or an environment---such that if the quantum state of the particle before the interaction is $\ket{s_1}$, then the quantum state of $E$ will become $\ket{E_1}$ (and similarly for $\ket{s_2}$), resulting in the final composite states $\ket{s_1}\ket{E_1}$ and $\ket{s_2}\ket{E_2}$, respectively. For an initial superposition state $\alpha\ket{s_1}+\beta\ket{s_2}$, the final composite state will be entangled,
\begin{equation}
\label{eq:1dlkf}
\ket{\Psi} = \alpha \ket{s_1} \ket{E_1} + \beta \ket{s_2} \ket{E_2}.
\end{equation}

The statistics of all possible local measurements on $S$ are exhaustively encoded in the reduced density matrix $\rho_S$,
\begin{align}
  \label{eq:aa12}
 \rho_S &= \text{Tr}_E(\rho_{SE}) =\text{Tr}_E \ketbra{\Psi}{\Psi} \nonumber \\
 &= \abs{\alpha}^2
    \ketbra{s_1}{s_1} + \abs{\beta}^2 \ketbra{s_2}{s_2} \nonumber \\ & \quad + \alpha \beta^*\ketbra{s_1}{s_2}
      \braket{E_2}{E_1} +  \alpha^*\beta\ketbra{s_2}{s_1}
      \braket{E_1}{E_2}.
\end{align}
For example, suppose we measure particle's position by letting the particle impinge on a distant detection screen. Statistically, the resulting particle probability density $p(x)$ will be given by
\begin{align}\label{eq:fdljksjk2}
p(x) & = \text{Tr}_S(\rho_S x) \notag \\ &= \abs{\alpha}^2
\abs{\psi_1(x)}^2 +\abs{\beta}^2 \abs{\psi_2(x)}^2 \notag \\ &\quad + 2 \,\text{Re} \left\{\alpha \beta^* \psi_1(x) \psi_2^*(x)\braket{E_2}{E_1} \right\},
\end{align}
where $\psi_i(x) \equiv \braket{x}{s_i}$. The last term represents the interference contribution. Thus, the visibility of the interference pattern is quantified by the overlap $\braket{E_2}{E_1}$, i.e., by the distinguishability of $\ket{E_1}$ and $\ket{E_2}$. In the limiting case of perfect distinguishability, $\braket{E_2}{E_1} = 0$, no interference pattern will be observable and we obtain the classical prediction. Phase relations have become \emph{locally} (i.e., with respect to $S$) inaccessible, and there is no measurement on $S$ that can reveal coherence between $\ket{s_1}$ and $\ket{s_2}$. The coherence is now between the states  $\ket{s_1} \ket{E_1}$ and $\ket{s_2} \ket{E_2}$, requiring an appropriate \emph{global} measurement (acting jointly on $S$ and $E$) for it to be revealed.

Conversely, if the interaction between $S$ and $E$ is such that $E$ is completely unable to resolve the path of the particle, then $\ket{E_1}$ and $\ket{E_2}$ are indistinguishable and full coherence is retained at the level of $S$, as is also directly obvious from Eq.~\eqref{eq:1dlkf}. In the intermediary regime where $0 < \abs{\braket{E_2}{E_1}} < 1$, meaning that $\ket{E_1}$ and $\ket{E_2}$ can be distinguished in a one-shot measurement with nonzero probability $p = 1-\abs{\braket{E_2}{E_1}}^2 < 1$, an interference pattern of reduced visibility is obtained. Equation~\eqref{eq:fdljksjk2} shows that the reduction in visibility increases as $\ket{E_1}$ and $\ket{E_2}$ become more distinguishable.

Here is another way of putting the matter. Looking back at Eq.~\eqref{eq:1dlkf}, we see that $E$ encodes which-way information about $S$ in the same ``relative-state'' sense \cite{Everett:1957:rw} in which EPR correlations \cite{Einstein:1935:dr,Bell:1964:ep,Bell:1966:ph} may be said to encode ``information.'' That is, if $\braket{E_2}{E_1} = 0$ and we were to measure $E$ and found it to be in state $\ket{E_1}$, we could, in EPR's words \cite[p.~777]{Einstein:1935:dr}, ``predict with certainty'' that we will find $S$ in $\ket{s_1}$.\footnote{Of course, this must not be read as saying that $S$ was already in $\ket{s_1}$ (i.e., ``went through slit 1'') prior to the measurement of $E$. Nor does it mean that the result of a subsequent path measurement on $S$ is necessarily determined, by virtue of the measurement on $E$, prior to this $S$-measurement's actually being carried out. After all, as Peres has cautioned us \cite{Peres:1978:aa}, unperformed measurements have no outcomes. So while the picture of $E$ as ``encoding which-path information'' about $S$ is certainly suggestive and helpful, it should be used with an understanding of its conceptual pitfalls.} Whenever such a prediction is possible were we to measure $E$, no interference effects between the components $\ket{s_1}$ and $\ket{s_2}$ can be measured at $S$, even if $E$ is never actually measured. If $\abs{\braket{E_2}{E_1}} > 0$, then $E$ encodes only \emph{partial} which-way information about $S$, in the sense that a measurement of $E$ could not reliably distinguish between $\ket{E_1}$ and $\ket{E_2}$; instead, sometimes the measurement will result in an outcome compatible with both $\ket{E_1}$ and $\ket{E_2}$. Consequently, an interference experiment carried out on $S$ would find reduced visibility, representing diminished local coherence between the components $\ket{s_1}$ and $\ket{s_2}$.

As hinted above, the description developed so far describes the essence of the decoherence process if we identify the particle $S$ more generally with an arbitrary quantum system and the second system $E$ with the environment of $S$. Then an idealized account of the decoherence interaction has form 
\begin{equation}
\label{eq:d1}
\biggl( \sum_i c_i \ket{s_i} \biggr) \ket{E_0} \quad \longrightarrow \quad \sum_i c_i \ket{s_i} \ket{E_i(t)}.
\end{equation}
We have here introduced a time parameter $t$, where $t=0$ corresponds to the onset of the environmental interaction, with $\ket{E_i(t)} \equiv \ket{E_0}$ for all $i$; at $t<0$ the system and environment are assumed to be uncorrelated (an assumption common to most decoherence models). 

A single environmental particle interacting with the system will typically only insufficiently resolve the components $\ket{s_i}$ in the system's superposition state. But because of the large number of such particles (and, hence, degrees of freedom), the overlap between their different joint states $\ket{E_i(t)}$ will rapidly decrease as a result of the build-up of many interaction events. Specifically, in many decoherence models an exponential decay of overlap is found \cite{Zurek:1982:tv,Joos:1985:iu,Paz:1993:ta,Leggett:1987:pm,Mokarzel:2002:za,Hornberger:2003:un,Zurek:2002:ii,Schlosshauer:2007:un,Breuer:2002:oq},
\begin{equation}
  \label{eq:jkHjkfhjhPJHyudf615}
  \braket{E_i(t)}{E_j(t)} \, \propto \, \E^{-t/\tau_\text{d}} \qquad \text{for $i \not= j$}.
\end{equation}
Here $\tau_\text{d}$ is the characteristic decoherence timescale, which can be evaluated for particular choices of the parameters in each model (see Sec.~\ref{sec:decmodels}).

\subsection{\label{sec:envir-monit-inform}Environmental monitoring and information transfer}

We will now motivate, in a different and more rigorous way, the picture of decoherence as a process of environmental monitoring. First, we express the influence of the environment in a completely general way. We assume that at $t=0$ there are no correlations between system $S$ and environment $E$, $\rho_{SE}(0) = \rho_S(0) \otimes \rho_E(0)$. We write $\rho_E(0)$ in its diagonal decomposition, $\rho_E(0) = \sum_i p_i \ketbra{E_i}{E_i}$, where $\sum_i p_i =1$ and the states $\ket{E_i}$ form an orthonormal basis of the Hilbert space of $E$. If $H$ denotes the Hamiltonian (here assumed to be time-independent) of $SE$ and $U(t) = \E^{-\I H t}$ represents the unitary time evolution operator, then the density matrix of $S$ evolves according to 
\begin{align}
  \label{eq:1slvjhvkjfkjvsj0}
  \rho_S(t) &= \text{Tr}_E \left\{ U(t) \left[ \rho_S(0) \otimes \left( \sum_i p_i \ketbra{E_i}{E_i} \right) \right] U^\dagger(t) \right\}\nonumber \\
&= \sum_{ij} p_i \bra{E_j} U(t) \ket{E_i} \rho_S(0)\bra{E_i} U^\dagger(t) \ket{E_j}.
\end{align}
Introducing the \emph{Kraus operators} \cite{Kraus:1983:ee} defined by $E_{ij} \equiv \sqrt{p_i} \bra{E_j} U(t) \ket{E_i}$, we obtain
\begin{equation}
 \rho_S(t) = \sum_{ij} E_{ij} \rho_S(0) E^\dagger_{ij}.
\end{equation}
It is customary to combine the two indices $i$ and $j$ into a single index and write the Kraus operators as
\begin{equation}
  \label{eq:worihfvsjvttrafs2}
  W_k \equiv \sqrt{p_{i_k}} \bra{E_{j_k}} U(t) \ket{E_{i_k}},
\end{equation}
such that 
\begin{equation}
\label{eq:dfjsb1}
 \rho_S(t) = \sum_k W_k \rho_S(0) W^\dagger_k.
\end{equation}
This Kraus-operator formalism (also called \emph{operator-sum formalism}) represents the effect of the environment as a sequence of (in general nonunitary) transformations of $\rho_S$ generated by the operators $W_k$. The Kraus operators exhaustively encode information about the initial state of the environment and about the dynamics of the joint $SE$ system. Because the evolution of $SE$ is unitary, the Kraus operators satisfy the completeness constraint
\begin{equation}
  \label{eq:19sirhgvksjvbkjsb}
\sum_k W_k W^\dagger_k = I_S,
\end{equation}
where $I_S$ is the identity operator in the Hilbert space of $S$. Equations~\eqref{eq:dfjsb1} and \eqref{eq:19sirhgvksjvbkjsb} together imply that the $W_k$ are the generators of a completely positive map $\Phi: \rho_S(0) \mapsto \rho_S$, also known as a \emph{quantum operation} \cite{Kraus:1983:ee} or \emph{quantum channel}.\footnote{The Kraus formalism is of limited use in calculating decoherence dynamics for concrete situations of physical interest. This is so because finding the Kraus operators corresponds to diagonalizing the full Hamiltonian of $SE$, usually a prohibitively difficult task. Moreover, the Kraus operators contain all contributions to the evolution of the reduced density matrix, while for considerations of decoherence we are typically interested only in the nonunitary terms, and certain contributions---such as back-action effects from the system on the environment---can often be neglected. (This is where master equations come into play; see Sec.~\ref{sec:mastereqs}.)} 

We will now use Eq.~\eqref{eq:dfjsb1} to formally motivate the view that decoherence corresponds to an indirect measurement of the system by the environment, and that it thus results from a transfer of information from the system to the environment (see also Ref.~\cite{Hornberger:2009:aq}). In such an indirect measurement, we let the system $S$ interact with a probe---here the environment $E$---followed by a projective measurement on $E$. The probe is treated as a quantum system. This procedure aims to yield information about $S$ without performing a projective (and thus destructive) direct measurement on $S$. To model such an indirect measurement, consider again an initial composite density operator $\rho_{SE}(0) = \rho_S(0) \otimes \rho_E(0)$ evolving under the action of $U(t) = \E^{-\I H t}$, where $H$ is the total Hamiltonian. Consider a projective measurement $M$ on $E$ with eigenvalues $\alpha$ and corresponding projectors $P_\alpha \equiv \ketbra{\alpha}{\alpha}$, with $P_\alpha^2=P_\alpha^\dagger=P_\alpha$. The probability of obtaining outcome $\alpha$ in this measurement when $S$ is described by the density operator $\rho_S(t)$ is
\begin{multline}
 \text{Prob}\left(\alpha \mid \rho_S(t) \right) = \text{Tr}_E \left( P_\alpha \rho_E(t) \right)  \\ =\text{Tr}_E \left\{ P_\alpha \text{Tr}_S \left[ U(t) \left( \rho_S(0) \otimes \rho_E(0) \right) U^\dagger(t)\right] \right\}.
\end{multline}
The density matrix of $S$ conditioned on the particular outcome $\alpha$ is
\begin{multline}
\rho_S^{(\alpha)}(t) = 
\frac{ \text{Tr}_E \left\{ \left[I \otimes P_\alpha \right] \rho_{SE}(t) \left[I \otimes P_\alpha \right] \right\}}{\text{Prob}\left(\alpha \mid \rho_S(t)\right)}\\
=\frac{ \text{Tr}_E \left\{ \left[I \otimes P_\alpha \right] U(t) \left[ \rho_S(0) \otimes \rho_E(0) \right] U^\dagger(t)  \left[I \otimes P_\alpha \right] \right\}}{\text{Prob}\left(\alpha \mid \rho_S(t)\right)}.\label{eq:hiuvb}
\end{multline}
Inserting the diagonal decomposition $\rho_E(0) = \sum_k p_k \ketbra{E_k}{E_k}$ and carrying out the trace gives \cite{Hornberger:2009:aq}
\begin{equation}
\rho_S^{(\alpha)}(t) = \sum_k \frac{ M_{\alpha,k}  \rho_S(t) M^\dagger_{\alpha,k}}{\text{Prob}\left(\alpha \mid \rho_S(t)\right)}.
\end{equation}
Here we have introduced the measurement operators
\begin{equation}
M_{\alpha,k} \equiv \sqrt{p_k} \bra{\alpha} U(t) \ket{E_k},
\end{equation}
which obey the completeness constraint $\sum_{\alpha,k}M_{\alpha,k}M_{\alpha,k}^\dagger=I_S$. Equation~\eqref{eq:hiuvb} describes the effect of the indirect measurement on the state of the system. If, however, we do not actually inquire about the result of this measurement, we must assign to the system a density operator that is a sum over all the possible conditional states $\rho_S^{(\alpha)}(t)$ weighted by their probabilities $\text{Prob}\left(\alpha \mid \rho_S(t)\right)$,
\begin{align}
\rho_S(t) &= \sum_\alpha \text{Prob}\left(\alpha \mid \rho_S(t)\right) \rho_S^{(\alpha)}(t) \notag\\&= \sum_{\alpha,k} M_{\alpha,k}  \rho_S(t) M^\dagger_{\alpha,k}.
\end{align}
Note that this expression is formally analogous to the Kraus-operator expression of Eq.~\eqref{eq:dfjsb1}, which described the effect of a general environmental interaction on the state of the system. Recall, further, that the situation we encounter in decoherence is precisely one in which we do not actually read out the environment---or, in the present picture, in which we do not inquire about the result of the indirect measurement. This suggests that decoherence can indeed be understood as an indirect measurement---a monitoring---of the system by its environment.

\subsection{\label{sec:envir-induc-supers}Environment-induced superselection and decoherence-free subspaces}

Decoherence can occur in any basis; which observable is monitored by the environment depends on the specific form of the system--environment interaction. The \emph{preferred states} (or \emph{preferred observables}) of the system emerge dynamically as those states that are the most robust to the interaction with the environment, in the sense that they become least entangled with the environment; thus, they are the states most immune to decoherence. This is the \emph{stability criterion} for the selection of preferred states, resulting in the dynamical selection of preferred states (``environment-induced superselection'') \cite{Zeh:1970:yt,Kubler:1973:ux,Zurek:1981:dd,Zurek:1982:tv}. These environment-superselected preferred states (or observables) are sometimes also called \emph{pointer states} (or \emph{pointer observables}) \cite{Zurek:1981:dd}, since they correspond to the physical quantities that are most easily ``read off'' at the level of the system, akin to the pointer on the dial of a measurement apparatus.

\subsubsection{Pointer states and the commutativity criterion}

To find the preferred states, we decompose the total system--environment Hamiltonian into the self-Hamiltonians of the system $S$ and environment $E$ representing the intrinsic dynamics, and a part $H_\text{int}$ representing the interaction between system and environment, 
\begin{equation}
H = H_S + H_E + H_\text{int}. 
\end{equation}
In many cases of practical interest, $H_\text{int}$ dominates the evolution of the system, $H \approx H_\text{int}$ (the \emph{quantum-measurement limit} of decoherence). We look for system states $\ket{s_i}$ such that the composite system--environment state, when starting from a product state $\ket{s_i}\ket{E_0}$ at $t=0$, remains in the product form $\ket{s_i}\ket{E_i(t)}$ for all $t>0$ under the action of $H_\text{int}$ (we shall assume here that $H_\text{int}$ is not explicitly time-dependent). That is, we demand that (setting $\hbar \equiv 1$ from here on)
\begin{equation}
  \label{eq:gxlknn98ygya24}
  \E^{-\I H_\text{int} t} \ket{s_i}\ket{E_0}=
  \lambda_i \ket{s_i}\E^{-\I H_\text{int} t} \ket{E_0} \equiv  \ket{s_i}\ket{E_i(t)}.
\end{equation}
Thus, the pointer states $\ket{s_i}$ are the eigenstates of the part of the interaction Hamiltonian $H_\text{int}$ pertaining to the Hilbert space of the system, with eigenvalues $\lambda_i$. These states will be stationary under $H_\text{int}$ \cite{Zurek:1981:dd}. It follows that the pointer observable defined by $O_S = \sum_i o_i \ketbra{s_i}{s_i}$ commutes with $H_\text{int}$,
\begin{equation}
  \label{eq:dhvvsdnbbfvs27}
  \bigl[ O_S, H_\text{int} \bigr] = 0.
\end{equation}
This \emph{commutativity criterion} \cite{Zurek:1981:dd,Zurek:1982:tv} is particularly easy to apply when $H_\text{int}$ takes the tensor-product form $H_\text{int} = S \otimes E$, as is frequently the case. Then the environment-superselected observables will be those observables that commute with $S$. If $S$ is Hermitian, it represents the physical quantity monitored by the environment. In general, any $H_\text{int}$ can be written as a diagonal decomposition of (unitary but not necessarily Hermitian) system and environment operators $S_\alpha$ and $E_\alpha$, $H_\text{int} =  \sum_\alpha S_\alpha \otimes E_\alpha$.
If the $S_\alpha$ are Hermitian, such a Hamiltonian represents the simultaneous environmental monitoring of different observables $S_\alpha$ of the system. A sufficient condition for $\{ \ket{s_i} \}$ to form a set of pointer states of the system is then given by the requirement that the $\ket{s_i}$ be simultaneous
eigenstates of the operators $S_\alpha$, 
\begin{equation}
  \label{eq:OIbvsrhjkbv9}
  S_\alpha \ket{s_i} = \lambda_i^{(\alpha)}\ket{s_i} \qquad
  \text{for all $\alpha$ and $i$}. 
\end{equation}

Interaction Hamiltonians frequently describe the scattering of surrounding particles (photons, air molecules, etc.), leading to \emph{collisional decoherence} (see Sec.~\ref{sec:collisionaldecoherence}). Since the force laws describing such processes typically depend on some power of distance, the interaction Hamiltonian will then commute with the position operator. Thus, the pointer states will be approximate eigenstates of position (i.e., narrow position-space wave packets). This explains why superpositions of mesoscopically and macroscopically distinct positions are prohibitively difficult to observe \cite{Zurek:1981:dd,Zurek:1982:tv,Joos:1985:iu,Zurek:1991:vv,Gallis:1990:un,Diosi:1995:um,Hornberger:2003:un,Hornberger:2006:tb,Hornberger:2008:ii,Busse:2009:aa,Busse:2010:aa}. Collisional decoherence can also be dominant in microscopic systems when these systems occur in distinct spatial configurations that couple strongly to the surrounding medium. For example, chiral molecules such as sugar are always observed to be in chirality eigenstates (left-handed or right-handed), which are superpositions of different energy eigenstates. Any attempt to prepare such molecules in energy eigenstates leads to immediate decoherence into the environmentally stable chirality eigenstates \cite{Harris:1981:rc,Zeh:1999:qr}.

The \emph{quantum limit of decoherence} \cite{Paz:1999:vv} arises when the modes of the environment are slow in comparison with the evolution of the system---that is, when the highest frequencies (i.e., energies) available in the environment are smaller than the separation between the energy eigenstates of the system. Then the environment will be able to monitor only quantities that are constants of motion. In the case of nondegeneracy, this quantity will be the energy of the system, leading to the environment-induced superselection of energy eigenstates for the system \cite{Paz:1999:vv}.\footnote{Textbooks on quantum mechanics usually attribute a special role to such energy eigenstates (for closed systems) since they are stationary under the action of the Hamiltonian. In this closed-system picture, however, arbitrary superpositions of energy eigenstates should nonetheless be perfectly legitimate. Thus, it is important to realize that the environment-induced superselection of energy eigenstates is \emph{not} equivalent to a situation in which the presence of the environment could be neglected altogether; instead, the environment plays the crucial role of continuously monitoring the energy of the system, leading to a local suppression of coherence between energy eigenstates.} 

In many realistic situations, the commutativity criterion, Eq.~\eqref{eq:dhvvsdnbbfvs27}, can only be fulfilled approximately \cite{Zurek:1993:qq,Zurek:1993:pu}. In addition, the self-Hamiltonian of the system and the interaction Hamiltonian may contribute in roughly equal strengths (e.g., in models for quantum Brownian motion  \cite{Hu:1992:om,Paz:2001:aa}; see Sec.~\ref{sec:quant-brown-moti}), rendering neither the quantum-measurement limit of negligible intrinsic dynamics nor the quantum limit of decoherence of a slow environment appropriate. In such cases, more general methods for determining the preferred states are required. The \emph{predictability-sieve strategy} \cite{Zurek:1993:pu,Zurek:1998:re,Zurek:1993:qq} computes the time dependence of the amount of decoherence introduced into the system for a large set of initial states of the system evolving under the total system--environment Hamiltonian. Typically, this decoherence is measured using either the purity $\text{Tr} \left(\rho_S^2 \right)$ or the von Neumann entropy $S(\rho_S) = - \text{Tr}\left( \rho_S \log_2 \rho_S \right)$ of the reduced density matrix $\rho_S$. The states most immune to decoherence will be those which lead to the smallest decrease in purity or the smallest increase in von Neumann entropy. Application of this method leads to a ranking of the possible preferred states with respect to their robustness to the interaction with the environment. For particular models it has been explicitly shown that the
states picked out by the predictability sieve are robust to the particular choice of the measure of decoherence. For example, in the model for quantum Brownian motion, different measures lead to the same minimum-uncertainty wave packets in phase space \cite{Kubler:1973:ux,Zurek:1993:pu,Zurek:2002:ii,Diosi:2000:yn,Joos:2003:jh,Eisert:2003:ib}.

\subsubsection{\label{sec:dfs}Decoherence-free subspaces}

The pointer-state condition of Eq.~\eqref{eq:OIbvsrhjkbv9} can be strengthened to the concept of \emph{pointer subspaces} \cite{Zurek:1982:tv} or \emph{decoherence-free subspaces} (DFS) \cite{Palma:1996:yy,Lidar:1998:uu,Zanardi:1997:yy,Zanardi:1997:tv,Zanardi:1998:oo,Lidar:1999:fa,%
  Bacon:2000:yy,Duan:1998:yb,Zanardi:2001:oo,Knill:2000:aa}. These are subspaces of the Hilbert space of the system in which \emph{every} state in the subspace is immune to decoherence; this is a nontrivial requirement, since in general superpositions of pointer states will not be pointer states themselves. One important condition for this to happen is that the preferred states $\ket{s_i}$ defined by Eq.~\eqref{eq:OIbvsrhjkbv9} form an orthonormal basis of the subspace, and that the eigenvalues $\lambda_i^{(\alpha)}$ in Eq.~\eqref{eq:OIbvsrhjkbv9} are independent of $i$, i.e., that all $\ket{s_i}$ are simultaneous \emph{degenerate} eigenstates of each $S_\alpha$,
\begin{equation}
  \label{eq:OIbvsrhjkbvsfljvh9}
  S_\alpha \ket{s_i} = \lambda^{(\alpha)} \ket{s_i} \qquad
  \text{for all $\alpha$ and $i$}. 
\end{equation}
This condition states that the action of a given $S_\alpha$ must be the same for all basis states $\ket{s_i}$ of the DFS, and thus the existence of a DFS corresponds to a symmetry in the structure of the system--environment interaction, i.e., to a \emph{dynamical symmetry}. A necessary condition for such a symmetry to obtain is the absence of terms in $H_\text{int}$ that act jointly on system and environment in a nontrivial manner.  

An arbitrary state $\ket{\psi}$ in the DFS can then be written as $\ket{\psi} = \sum_i c_i \ket{s_i}$ and will evolve according to
\begin{align}
  \label{eq:OIbvsrhjkbv9zFFHGSVCxc}
  \E^{-\I H_\text{int}t} \ket{\psi}\ket{E_0} &= \ket{\psi}\E^{-\I
    \left( \sum_\alpha \lambda^{(\alpha)} E_\alpha \right)t}
  \ket{E_0} \notag\\&\equiv \ket{\psi} \ket{E_\psi(t)}.
\end{align}
Thus, the state $\ket{\psi}$ does not become entangled with the environment and is therefore immune to decoherence. When the self-Hamiltonian $H_S$ of the system cannot be neglected, one needs to additionally ensure that none of the basis states $\ket{s_i}$ of the DFS will drift out of the subspace under the evolution generated by $H_S$. Otherwise an initially decoherence-free state would again become prone to decoherence. The concept of DFS can be generalized to the formalism of \emph{noiseless subsystems} (or \emph{noiseless quantum codes}) \cite{Knill:2000:aa,Kempe:2001:oo,Lidar:2003:aa}. 

\subsection{\label{sec:prol-inform-quant}Proliferation of information and quantum Darwinism}

\emph{Quantum Darwinism} \cite{Zurek:2003:pl,Ollivier:2003:za,Ollivier:2004:im,Blume:2004:oo,Blume:2005:oo,Zurek:2009:om,Riedel:2010:un,Riedel:2011:un,Riedel:2012:un} builds on the ideas of decoherence and environmental encoding of information, by broadening the role of the environment to that of a communication and amplification channel. Interactions between the system and its environment lead to the redundant storage of selected information about the system in  many fragments of the environment. By measuring some of these fragments, observers can indirectly obtain information about the system without appreciably disturbing the system itself. Indeed, this represents how we typically observe objects. For example, we see an object not by directly interacting with it, but by intercepting scattered photons that encode information about the object's spatial structure \cite{Riedel:2010:un,Riedel:2011:un}. 

In this sense, quantum Darwinism provides a dynamical explanation for the robustness of states of (especially) macroscopic objects to observation. It was found that the observable of the system that can be imprinted most completely and redundantly in many distinct fragments of the environment coincides with the pointer observable selected by the system--environment interaction  \cite{Ollivier:2003:za,Ollivier:2004:im,Blume:2004:oo,Blume:2005:oo}; conversely, most other states do not seem to be redundantly storable. Indeed, it has been shown that the redundant proliferation of information regarding pointer states is as inevitable as decoherence itself \cite{Zwolak:2014:tt}. Quantum Darwinism has been studied in several concrete models, for example, in spin environments \cite{Blume:2004:oo}, quantum Brownian motion \cite{Blume:2007:oo}, and photon and photon-like environments \cite{Riedel:2010:un,Riedel:2011:un,Zwolak:2014:tt}. The efficiency of the amplification process described by quantum Darwinism can be expressed in terms of the \emph{quantum Chernoff information} \cite{Zwolak:2014:tt}.

The structure and amount of information that the environment encodes about the system can be quantified using the measure of (classical \cite{Ollivier:2003:za,Ollivier:2004:im} or quantum \cite{Zurek:2002:ii,Blume:2004:oo,Blume:2005:oo}) \emph{mutual information}. Classical mutual information is based on the choice of particular observables of the system $S$ and the environment $E$ and quantifies how well one can predict the outcome of a measurement of a given observable of $S$ by measuring some observable on a fraction of $E$ \cite{Ollivier:2003:za,Ollivier:2004:im}. Quantum mutual information is defined as $S(\rho_S) + S(\rho_E) - S(\rho)$, where $\rho_S$, $\rho_E$, and $\rho$ are the density matrices of $S$, $E$, and the composite system $SE$, respectively, and $S(\rho) = - \text{Tr}\left( \rho \log_2 \rho \right)$ is the von Neumann entropy associated with $\rho$. Quantum mutual information quantifies the degree of quantum correlations between $S$ and $E$. Classical and quantum mutual information give similar results \cite{Ollivier:2003:za,Ollivier:2004:im,Zurek:2002:ii,Blume:2004:oo,Blume:2005:oo} because the difference between the two measures, known as the \emph{quantum discord} \cite{Ollivier:2001:az}, disappears when decoherence is sufficiently effective to select a well-defined pointer basis \cite{Ollivier:2001:az}. 

\subsection{\label{sec:decoh-vers-diss}Decoherence versus dissipation and noise}

While the presence of dissipation implies the presence of decoherence, the converse is not necessarily true. When dissipation and decoherence are both present, they typically occur on vastly different timescales; the decoherence timescale is typically many orders of magnitude shorter than the relaxation timescale. A rule-of-thumb estimate for the ratio of the relaxation timescale $\tau_\text{r}$ to the decoherence timescale $\tau_\text{d}$ for a massive object described by a superposition of two different positions a distance $\Delta x$ apart is \cite{Zurek:1986:uz}
\begin{equation}
  \label{eq:daf12}
  \frac{\tau_\text{r}}{\tau_\text{d}} \sim \left( \frac{\Delta
      x}{\lambda_\text{dB}} \right)^2,
\end{equation}
where $\lambda_\text{dB}=(2mkT)^{-1/2}$ is the thermal de Broglie wavelength of the object. For an object of mass $m = \unit[1]{g}$ at room temperature in a coherent superposition of two locations a distance $\Delta x =\unit[1]{cm}$ apart, $\tau_\text{r}/\tau_\text{d}$ is on the order of $10^{40}$. Thus, for macroscopic objects the dissipative influence of the environment is usually completely negligible on the timescale relevant to the decoherence induced by this environment. 

Decoherence is a consequence of environmental entanglement. In the literature on quantum computing, however, the term ``decoherence'' is often used to refer to \emph{any} process that affects the qubits, including perturbations due to \emph{classical} fluctuations and imperfections. Examples for sources of such classical noise in the context of quantum computing are the fluctuations in the intensity \cite{Schneider:1998:yz} and duration \cite{Miquel:1997:zz} of the laser beam incident on qubits in an ion trap, inhomogeneities in the magnetic fields in NMR quantum computing \cite{Vandersypen:2004:ra}, and bias fluctuations in superconducting qubits \cite{Martinis:2003:bz}. The distinction between classical noise and quantum decoherence has been further blurred in quantum error correction, since the error-correcting schemes are insensitive to the physical origin of the qubit errors (see Sec.~\ref{sec:corr-decoh-induc}).

Phenomenologically and formally the influence of classical noise processes may be described in a manner similar to the effect of environmental entanglement, namely, in terms of a decay of the off-diagonal elements (interference terms) in the local density matrix (in the environment-superselected basis). But in the case of noise, the decay of the off-diagonal elements occurs because the system's density matrix is identified with an average over a physical ensemble of systems (or, put differently, over the different instances of particular noise processes), while in the case of decoherence the decay is due to an entanglement-induced delocalization of phase coherence for individual systems. The fundamental difference between these physical processes is masked by the density-matrix description. Indeed, one can always find an experimental procedure that would, at least in principle, distinguish between the different physical processes underlying formally similar density-matrix descriptions.

In contrast with decoherence, noise does not create system--environment entanglement and can in principle always be undone using only local operations (witness, for example, the reversal of ensemble dephasing in NMR experiments using the spin-echo technique). In any individual realization of the noise process the dynamics of the system are completely unitary, and thus no coherence is lost from the system. By contrast, if the system becomes entangled with environmental degrees of freedom, at the very least we would need to perform a pair of measurements on the environment before and after the interaction with the system in order to gather enough information to reverse the effect of decoherence by application of an appropriate countertransformation. Moreover, as also seen experimentally \cite{Myatt:2000:yy}, these measurements would not always constitute a sufficient procedure for ``undoing'' decoherence (see also Sec.~IV.C of Ref.~\cite{Zurek:2002:ii}). 

The loss of phase coherence due to environmental entanglement is sometimes \emph{simulated} (with the above caveats) by classical fluctuations perturbing the system, i.e., by the addition of certain time-dependent terms to the self-Hamiltonian of the system. This strategy was implemented, for example, in theoretical \cite{Schneider:1998:yz,Schneider:1999:tt} and experimental \cite{Turchette:2000:aa,Myatt:2000:yy} studies of the influence of fluctuating parameters in ion-trap quantum computers.

\section{\label{sec:mastereqs}Master equations}

In the usual approach to modeling decoherence, the reduced density matrix $\rho_S(t)$ is obtained from
\begin{equation}
\label{eq:dmm2}
  \rho_S(t) = \text{Tr}_E \, \rho_{SE}(t) \equiv \text{Tr}_{E} \left\{ U(t) \rho_{SE}(0) U^\dagger(t) \right\},
\end{equation}
where $U(t)$ is the time-evolution operator for the composite system $SE$. The task of calculating $\rho_{SE}(t)$ is often computationally cumbersome or even intractable. It is also unnecessarily detailed, because we are usually only interested in the dynamics of the system. A \emph{master equation} allows us to calculate $\rho_S(t)$ directly from an expression of the form
\begin{equation}\label{eq:dmm}
\rho_S(t) = \mathcal{V}(t) \rho_S(0),
\end{equation}
where the superoperator $\mathcal{V}(t)$ is the \emph{dynamical map} generating the evolution of $\rho_S(t)$. If the master equation is exact, then we merely have the identity $\mathcal{V}(t) \rho_S(0) \equiv \text{Tr}_{E} \left\{ U(t) \rho_{SE}(0) U^\dagger(t) \right\}$ and no computational advantage is gained. Therefore, master equations are typically based on simplifying approximations. 

In modeling decoherence, we focus on master equations that are first-order time-local differential equations of the form
\begin{equation}
\label{eq:5vshvrgyde1}
\frac{\D}{\D t} \rho_S(t) = \mathcal{L}\left[\rho_S(t)\right] \equiv -\I  \left[ H'_S, \rho_S(t) \right] + \mathcal{D}[\rho_S(t)].
\end{equation}
This equation is local in time in the sense that the change of $\rho_S$ at time $t$ depends only on $\rho_S$ evaluated at $t$. The superoperator $\mathcal{L}$ acting on $\rho_S(t)$ typically depends on the initial state of the environment and the different terms in the Hamiltonian. We have decomposed $\mathcal{L}$ into two parts to distinguish their physical interpretation. The first term, $-\I  \left[ H'_S, \rho_S(t) \right]$, is unitary and given by the Liouville--von Neumann commutator with the ``renormalized'' Hamiltonian $H'_S$ of the system. (Because the environment typically leads to a renormalization of the energy levels of the system, this Hamiltonian does in general not coincide with the unperturbed free Hamiltonian $H_S$ of $S$ that would generate the evolution of $S$ in absence of the environment.) The second, nonunitary term $\mathcal{D}[\rho_S(t)]$ represents decoherence (and often also dissipation) due to the environment.

\subsection{Born--Markov master equations}

Born--Markov master equations allow for many decoherence problems to be treated in a mathematically simple, and often closed, form. They are based on the following two approximations:

\begin{enumerate}

\item \emph{The Born approximation.} The system--environment coupling is sufficiently weak and the environment is reasonably large such that changes of the density operator $\rho_E$ of the environment are negligible and the system--environment density operator remains remains approximately factorized at all times, $\rho_{SE}(t) \approx \rho_S(t) \otimes \rho_E$.

\item \emph{The Markov approximation.}  Memory effects of the environment are negligible, in the sense that any self-correlations within the environment created
  by the coupling to the system decay rapidly compared to the characteristic relaxation timescale of the open quantum system. 

\end{enumerate}

Comparisons between the predictions of models based on Born--Markov master equations and experimental data indicate that the Born and Markov assumptions are reasonable in many physical situations (but see Sec.~\ref{sec:non-mark-decoh} below for exceptions and non-Markovian models). Assuming these assumptions hold and writing the interaction Hamiltonian as $H_\text{int} = \sum_\alpha S_\alpha \otimes E_\alpha$, the Born--Markov master equation reads \cite{Breuer:2002:oq,Schlosshauer:2007:un}
\begin{multline}
\label{eq:born-markov-master}
\frac{\D}{\D t} \rho_S(t) = -\I \left[ H_S, \rho_S(t) \right] \\- \sum_{\alpha} \left\{ \left[
S_\alpha, B_\alpha \rho_S(t) \right] + \left[ \rho_S(t) C_\alpha, S_\alpha \right] \right\},
\end{multline}
where the system operators $B_\alpha$ and $C_\alpha$ are defined as
\begin{subequations} \label{eq:hvdwg643r5gsxkjcvbsvnx20}
\begin{align}
\label{eq:iuwrgf8}
B_\alpha &\equiv \int_0^\infty \D \tau \, \sum_{\beta}
c_{\alpha\beta}(\tau)  S^{(I)}_\beta(-\tau),
\\ \label{eq:112823jmn2} C_\alpha &\equiv \int_0^\infty \D \tau \,
\sum_{\beta} c_{\beta\alpha}(-\tau)  S^{(I)}_\beta(-\tau).
\end{align}
\end{subequations}
Here $S^{(I)}_\alpha(-\tau)$ denotes the operator $S_\alpha$ in the interaction picture. In the following, we will simplify notation by omitting the
superscript ``${I}$''; instead we use the convention that all operators bearing explicit time arguments are to be understood as interaction-picture operators. (For density operators, however, we will maintain the superscript notation in order to distinguish them from Schr\"odinger-picture density operators, which also carry a
time argument.) The quantities $c_{\alpha\beta}(\tau)$ appearing in Eq.~\eqref{eq:hvdwg643r5gsxkjcvbsvnx20} are given by
\begin{equation}
  \label{eq:pmpnnun12}
c_{\alpha\beta}(\tau) \equiv \left\langle E_\alpha(\tau) E_\beta
  \right\rangle_{\rho_E}.
\end{equation}
These \emph{environment self-correlation functions} quantify how much information the environment retains over time about its interaction with the system. The Markov approximation corresponds to the assumption of a rapid decay of the $c_{\alpha\beta}(\tau)$ relative to the timescale set by the evolution of the
system.

In many situations of interest, the general form of the Born--Markov master equation, Eq.~\eqref{eq:born-markov-master}, simplifies considerably. For example, typically only a single system observable $S$ is monitored by the environment, $H_\text{int} = S \otimes E$. Also, the time dependence of the operators $S_\alpha(\tau)$ and $E_\alpha(\tau)$ is often simple, facilitating the calculation of the quantities $B_\alpha$ and $C_\alpha$. Examples are discussed in Sec.~\ref{sec:decmodels}.

\subsection{Lindblad master equations}

Lindblad master equations constitute a particular, albeit quite general, class of time-local Markovian master equations. They arise from the requirement that the evolution of the reduced density matrix generated by the master equation must ensure  complete positivity \cite{Kraus:1971:ii,Gorini:1976:tt,Lindblad:1976:um,Breuer:2002:oq,Alicki:2001:aa,Alicki:2007:uu,Benatti:2005:ii}. Complete positivity guarantees that the dynamical map $\rho_{S}(0) \mapsto \rho_S(t) = \mathcal{V}(t) \rho_{S}(0)$ described by the master equation generates physically consistent dynamics even when $S$ is initially entangled with another system. While complete positivity is automatically fulfilled if the evolution is exact, approximate master equations will not necessarily ensure complete positivity \cite{Dumke:1979:ia,Breuer:2002:oq,Alicki:2001:aa,Alicki:2007:uu,Benatti:2005:ii}. The Lindblad master equation is a special case of the general Born--Markov master equation that ensures complete positivity and takes the general form \cite{Gorini:1976:tt,Lindblad:1976:um}
\begin{multline}
  \label{eq:sdfkhwr69}
\frac{\D}{\D t} \rho_S(t) = -\I \left[ H'_S, \rho_S(t) \right] \\+ \frac{1}{2} \sum_{\alpha\beta} \gamma_{\alpha\beta} \left\{ \left[ S_\alpha, \rho_S(t) S^\dagger_\beta
    \right] +  \left[ S_\alpha \rho_S(t), S^\dagger_\beta \right] \right\},
\end{multline}
where $H'_S$ is the renormalized Hamiltonian of the system. The coefficients $\gamma_{\alpha\beta}$ are time-independent and exhaustively encapsulate information about the physical parameters of the decoherence processes (and possibly dissipation processes). One can show that the matrix $\Gamma \equiv ( \gamma_{\alpha\beta} )$ formed by the coefficients $\gamma_{\alpha\beta}$ is positive, i.e., all its eigenvalues $\kappa_\mu$ are $\ge 0$. Therefore, Eq.~\eqref{eq:sdfkhwr69} can be simplified by diagonalizing $\Gamma$, which results in the diagonal form \cite{Lindblad:1976:um,Gorini:1978:uf}
\begin{multline}\label{eq:lindblad}
  \frac{\D}{\D t} \rho_S(t) = - \I \left[ H'_S, \rho_S(t) \right]  \\- \frac{1}{2} \sum_\mu \kappa_\mu \left\{ L_\mu^\dagger
    L_\mu \rho_S(t) + \rho_S L_\mu^\dagger L_\mu - 2L_\mu \rho_S(t) L_\mu^\dagger \right\}.
\end{multline}
The \emph{Lindblad operators} $L_\mu$ are linear combinations of the original operators $S_\alpha$, with coefficients determined by the diagonalization of $\Gamma$. The Lindblad structure of a master equation can also be motivated from the requirement that it gives rise to the most general form of generators of quantum dynamical semigroups \cite{Lindblad:1976:um,Gorini:1976:tt,Gorini:1978:uf,Davies:1974:tw,Kossakowski:1972:tf, Breuer:2002:oq,Alicki:2007:uu}. It is possible to bring any Born--Markov master equation into Lindblad form by imposing the \emph{rotating-wave approximation}. This assumption, ubiquituous in quantum optics, is justified whenever the timescale set by the typical energy differences $\hbar(\omega-\omega')$ of the system Hamiltonian is short in comparison with the relaxation timescale of the system. (See Sec.~3.3.1 of Ref.~\cite{Breuer:2002:oq} for details.) 

Because the $S_\alpha$ are not necessarily Hermitian, the Lindblad operators do not always correspond to physical observables. But when they do, we can rewrite Eq.~\eqref{eq:lindblad} in compact double-commutator form,
\begin{equation}\label{eq:lindbladc}
\frac{\D}{\D t} \rho_S(t) = - \I \left[ H'_S, \rho_S(t) \right] - \frac{1}{2} \sum_\mu \kappa_\mu \left[ L_\mu, \left[ L_\mu, \rho_S(t) \right]
\right].
\end{equation}
As an example, consider a situation in which the environment monitors the position of a system. With $L = x$ and the ``free''-particle Hamiltonian $H'_S = H_S = p^2/2m$, Eq.~\eqref{eq:lindbladc} becomes
\begin{equation}\label{eq:lifsfdndbladc}
\frac{\D}{\D t} \rho_S(t) =  -\frac{\I}{2m}\left[p^2, \rho_S(t) \right] - \frac{1}{2}  \kappa \left[ x, \left[ x, \rho_S(t) \right]\right].
\end{equation}
Expressing this master equation in the position representation results in
\begin{multline}\label{eq:lifsshvgvvvxayhcgiefdndbladc}
  \frac{\partial \rho_S(x,x',t)}{\partial t} = - \frac{\I}{2m} \left(\frac{ \partial^2}{\partial x'^2}- \frac{ \partial^2}{\partial x^2} \right) \rho_S(x,x',t)\\ -  \frac{1}{2}  \kappa
  \left(x-x'\right)^2   \rho_S(x,x',t).
\end{multline}
This is the classic equation of motion for decoherence due to environmental scattering first derived in Ref.~\cite{Joos:1985:iu}.

Lindblad master equations provide an intuitive and simple way of representing the environmental monitoring of an open quantum system. Most of the real physics behind this monitoring process is hidden in the coefficients $\kappa_\mu$ appearing in Eq.~\eqref{eq:lindblad}. If the Lindblad operators are chosen to be dimensionless, the $\kappa_\mu$ can be directly interpreted as decoherence rates, since they have units of inverse time. 

Equation \eqref{eq:lindbladc} shows that the decoherence term vanishes if 
\begin{equation}
  \label{eq:7fskjhytw10}
  \left[ L_\mu, \rho_S(t) \right] = 0 \qquad \text{for all $\mu,t$}.
\end{equation}
In this case, $\rho_S(t)$ evolves unitarily. Since the $L_\mu$ are linear combinations of the $S_\alpha$, Eq.~\eqref{eq:7fskjhytw10} typically means that
$\left[ S_\alpha, \rho_S(t) \right] = 0$ for all $\alpha, t$. This implies that simultaneous eigenstates of all $S_\alpha$ will be immune to decoherence, which is precisely the pointer-state criterion of Eq.~\eqref{eq:OIbvsrhjkbv9}.

In \emph{quantum-jump} and \emph{quantum-trajectory} approaches, the evolution of the reduced density matrix is conditioned on an explicitly observed sequence of measurement results in the environment. This allows for the (formal) description of a single realization of the system evolving stochastically, conditioned on a particular measurement record. The dynamics are then described by a master equation of the Lindblad type, Eq.~\eqref{eq:lindbladc}, for the reduced density matrix $\rho^C_S$ conditioned on the measurement records of the Lindblad operators $L_\mu$,
\begin{multline} \label{eq:cme}
\D \rho^C_S = -\I \left[H_S, \rho_S^C \right] \D t  - \frac{1}{2} \sum_\mu \kappa_\mu \left[L_\mu, \left[L_\mu, \rho_S^C\right] \right] \D t \\+ \sum_\mu \sqrt{\kappa_\mu} \, \mathcal{W}[L_\mu] \rho_S^C \, \D W_\mu.  
\end{multline}
Here, $\mathcal{W}[L]\rho \equiv L \rho + \rho L^\dagger - \rho \, \text{Tr} \left\{ L\rho + \rho L^\dagger \right\}$, and the $\D W_\mu$ denote so-called Wiener increments. Equation~\eqref{eq:cme} corresponds to a diffusive unraveling of the Lindblad equation into individual quantum trajectories, which can then be expressed by means of a \emph{stochastic Schr\"odinger equation} \cite{Barchielli:1991:fv,Belavkin:1989:an,Belavkin:1989:am,Belavkin:1989:um,Belavkin:1995:tt,Diosi:1988:wx,Diosi:1988:hn,Diosi:1988:bv,Gisin:1984:qs,Gisin:1989:jn,Wiseman:1994:qq,Goan:2001:rz,Plenio:1998:bb}.

\subsection{\label{sec:non-mark-decoh}Non-Markovian decoherence}

The derivation of the Born--Markov master equation assumes that the coupling between system and environment is weak and memory effects of the environment can be neglected. These conditions, however, are not met in certain situations of physical interest. An example would be a superconducting qubit strongly coupled to a low-temperature environment of other two-level systems \cite{Prokofev:2000:zz,Dube:2001:zz}. Also, a recent experiment \cite{Groeblacher:2013:im} has measured strongly non-Ohmic spectral densities for the environment of a quantum nanomechanical system; such densities lead to non-Markovian evolution. 

In many cases, pronounced memory effects in the environment will cause strong dependencies of the evolution of the reduced density operator on the past history of the system--environment composite and therefore make it impossible to describe the reduced dynamics by a differential equation that is local in time. Surprisingly, however, one can show that even non-Markovian dynamics sometimes can still be described by a time-local differential equation of the form
\begin{equation}
\label{eq:sfihvsfhv7}
  \frac{\D}{\D t} \rho_S(t) = \mathcal{K}(t) \rho_S(t),
\end{equation}
where the superoperator $\mathcal{K}(t)$ depends only on $t$. For example, a non-Markovian master equation for quantum Brownian motion (see Sec.~\ref{sec:quant-brown-moti}) can be obtained through a formal modification of the Born--Markov master equation \cite{Paz:2001:aa,Zurek:2002:ii}. In general, it is often possible to arrive at non-Markovian but time-local master equations via the so-called time-convolutionless projection operator technique \cite{Chaturvedi:1979:pm,Shibata:1980:ma,Royer:1972:um,Royer:2003:za}.

\section{\label{sec:decmodels}Decoherence models}

Many physical systems can be represented either by a qubit if the state space of the system is
discrete and effectively two-dimensional, or by a particle described by continuous phase-space coordinates. Needless to say, in the case of quantum information processing the qubit representation is of particular relevance. 

Similarly, a wide range of environments can be modeled as a collection of quantum harmonic oscillators or qubits. Harmonic-oscillator environments are of great generality. At low energies, many systems interacting with an environment can effectively be represented by one or two coordinates of the system linearly coupled to an environment of harmonic oscillators; indeed, sufficiently weak interactions with an arbitrary environment can be mapped onto a system linearly coupled to a harmonic-oscillator environment \cite{Feynman:1963:jj,Caldeira:1983:gv}.

Environments represented by qubits are often the appropriate model in the low-temperature regine, where decoherence is typically dominated by interactions with localized modes, such as paramagnetic spins, paramagnetic electronic impurities, tunneling charges, defects, and nuclear spins \cite{Dube:2001:zz,Prokofev:2000:zz,Lounasmaa:1974:yb}. Each of the localized modes is represented by a finite-dimensional Hilbert space with a finite energy cutoff. We can therefore model these modes as a set of discrete states. Typically, only two such states are relevant, and thus the localized modes can be mapped onto an environment of qubits. Since qubits can be formally represented by spin-$\frac{1}{2}$ particles, such models are known as \emph{spin-environment models}. 

In the following, we will discuss four important standard models, namely, collisional decoherence (Sec.~\ref{sec:collisionaldecoherence}), quantum Brownian motion (Sec.~\ref{sec:quant-brown-moti}), the spin--boson model (Sec.~\ref{sec:spin-boson-models}), and the spin--spin model (Sec.~\ref{sec:spin-envir-models}). For details on these and other decoherence models, including derivations of the relevant master equations, see Secs.~3 and 5 of Ref.~\cite{Schlosshauer:2007:un}. 

\subsection{\label{sec:collisionaldecoherence}Collisional decoherence}

Collisional decoherence arises from the scattering of environmental particles by a massive free quantum particle. Models of collisional decoherence were first studied in the classic paper by Joos and Zeh \cite{Joos:1985:iu}. A more rigorous and general treatment was later developed by Hornberger and collaborators \cite{Hornberger:2003:un,Hornberger:2006:tb,Hornberger:2008:ii,Busse:2009:aa,Busse:2010:aa} (see also \cite{Gallis:1990:un,Diosi:1995:um,Adler:2006:yb}), which, among other refinements, remedied a flaw in Joos and Zeh's original derivation that had resulted in decoherence rates that were too large by a factor of $2\pi$ \cite{Hornberger:2003:un}. 

If we assume that the central particle is much more massive than the environmental particles such that its center-of-mass state is not disturbed by the
  scattering events (no recoil), the time evolution of the reduced density matrix is given by \cite{Joos:1985:iu,Gallis:1990:un,Diosi:1995:um,Hornberger:2003:un,Schlosshauer:2007:un}
\begin{equation} 
\label{eq:scatq} 
\frac{\partial \rho_S(\bvec{x}, \bvec{x}', t)}{\partial t} 
= - F(\bvec{x} - \bvec{x}') \rho_S(\bvec{x}, \bvec{x}', t).
\end{equation}
This master equation describes pure spatial decoherence without dissipation. The decoherence factor $F(\bvec{x} - \bvec{x}')$ plays the role of a localization rate. It represents the characteristic decoherence rate at which spatial coherences between two positions $\bvec{x}$ and $\bvec{x}'$ become locally suppressed and is given by
\begin{multline}
\label{eq:scatf}
  F(\bvec{x} - \bvec{x}') =  \int_0^\infty \D q \,
 \varrho(q) v(q) \\\times \int \frac{\D \hat{n}\,\D \hat{n}'}{4\pi}  \left(1- \E^{\I
    q\left(\buvec{n} - \buvec{n}'\right) \cdot \left( \bvec{x} - \bvec{x'}
    \right) } \right) \abs{ f(q\buvec{n}, q\buvec{n}') }^2.
\end{multline}
Here $\varrho(q)$ denotes the number density of incoming particles with magnitude of momentum equal to $q=\abs{\bvec{q}}$, $\buvec{n}$ and $\buvec{n}'$ are unit vectors (with $\D \hat{n}$ and $\D \hat{n}'$ representing the associated solid-angle differentials), and $v(q)$ denotes the speed of particles with momentum $q$. For the scattering of massive environmental particles we have $v(q) = q/m$, where $m$ is each particle's mass, while for the scattering of photons and other massless particles $v(q)$ is equal to the speed of light. The quantity $\abs{ f(q\buvec{n}, q\buvec{n}')}^2$ is the differential cross section for the scattering of an environmental particle from initial momentum $\bvec{q}=q\buvec{n}$ to final momentum $\bvec{q}'=q\buvec{n}'$.

Whenever the mass of the central particle becomes comparable to the mass of the environmental particles (as in the case of air molecules scattered by small molecules and free electrons \cite{Tegmark:1993:uz}), the no-recoil assumption does not hold and more general models for collisional decoherence have to be considered \cite{Diosi:1995:um,Hornberger:2006:tb}. The resulting dynamics include dissipation, as well as decoherence in both position and momentum. 

To further evaluate the decoherence factor $F(\bvec{x} - \bvec{x}')$, Eq.~\eqref{eq:scatf}, we distinguish two important limiting cases. In the \emph{short-wavelength limit}, the typical wavelength of the scattered environmental particles is much shorter than the
coherent separation $\Delta x = \abs{\bvec{x}-\bvec{x}'}$ between the well-localized wave packets in the spatial superposition state of the system. Then a single scattering event will be able to fully resolve this separation and thus carry away complete which-path information, leading to maximum spatial decoherence per scattering event. In this limit, $F(\bvec{x} - \bvec{x}')$ turns out to be simply equal to the total scattering rate $\Gamma_\text{tot}$ \cite{Schlosshauer:2007:un}. This implies the existence of an upper limit for the decoherence rate when increasing the separation $\Delta x$, in contrast with decoherence rates obtained from linear models [compare Eqs.~\eqref{eq:daf12} and \eqref{eq:odijsvuhfsw21}]. Equation~\eqref{eq:scatq} then shows that spatial interference terms will become exponentially suppressed at a rate set by $\Gamma_\text{tot}$,
\begin{equation}\label{eq:sees}
\rho_S(\bvec{x},\bvec{x}',t) =
\rho_S(\bvec{x},\bvec{x}',0) \E^{-\Gamma_\text{tot} t}.
\end{equation}
In the opposite \emph{long-wavelength limit}, the environmental wavelengths are much larger than the coherent separation $\Delta x = \abs{\bvec{x}-\bvec{x}'}$, which implies that an individual scattering event will reveal only incomplete which-path information. For this case, one can show that spatial coherences become exponentially suppressed at a rate that depends on the square of the separation $\Delta x$ \cite{Schlosshauer:2007:un},
\begin{equation}\label{eq:scwer2}
\rho_S(\bvec{x},\bvec{x}',t) =
\rho_S(\bvec{x},\bvec{x}',0) \E^{-\Lambda (\Delta x)^2 t},
\end{equation}
where $\Lambda$ is a \emph{scattering constant} that encapsulates the physical details of the interaction. Thus, the quantity $\Lambda (\Delta x)^2$ plays the role of a decoherence rate. The dependence of this rate on $\Delta x$ is reasonable: if the environmental wavelengths are much larger than $\Delta x$, it will require a large number of scattering events to encode an appreciable amount of which-path information in the environment, and this amount will increase, for a given number of scattering events, as $\Delta x$ becomes larger. Note that if $\Delta x$ is increased beyond the typical wavelength of the environment, the short-wavelength limit needs to be considered instead, for which the decoherence rate is independent of $\Delta x$ and attains its maximum possible value.

Numerical values of collisional decoherence rates obtained from Eqs.~\eqref{eq:sees} and \eqref{eq:scwer2}, with the physically relevant scattering parameters $\Gamma_\text{tot}$ and $\Lambda$ appropriately evaluated, have shown the extreme efficiency of collisions in suppressing spatial interferences; Table~\ref{tab:decrate} shows a few classic order-of-magnitude estimates \cite{Joos:1985:iu,Joos:2003:jh,Schlosshauer:2007:un}. Excellent agreement between theory and experiment has been demonstrated for the decoherence of fullerenes due to collisions with background gas molecules in a Talbot--Lau interferometer \cite{Hackermuller:2003:uu,Hornberger:2003:tv,Hornberger:2003:un,Hornberger:2004:bb,Nimmrichter:2011:pr} (see Sec.~\ref{sec:matt-wave-interf} and Fig.~\ref{fig:c70-vis}), and for the decoherence of sodium atoms in a Mach--Zehnder interferometer due to the scattering of photons \cite{Kokorowski:2001:ub} and gas molecules \cite{Uys:2005:yb}.

\begin{table}[t]
\centering
\caption{Estimates of decoherence timescales (in seconds) for the
  suppression of spatial interferences over a distance $\Delta x$
  equal to the size $a$ of the object ($\Delta x = a = \unit[10^{-3}]{cm}$ for
  a dust grain and $\Delta x = a = \unit[10^{-6}]{cm}$ for a large
  molecule). See Ref.~\cite{Schlosshauer:2007:un} for details.}
\label{tab:decrate}
\begin{tabular}{lcc}
  \hline\noalign{\smallskip}
  Environment & \,\,Dust grain\,\, &  Large molecule \\
  \noalign{\smallskip}\hline\noalign{\smallskip}
  Cosmic background radiation  & $1$  & $10^{24}$ \\
  Photons at room temperature  & $10^{-18}$  & $10^{6}$ \\
  Best laboratory vacuum  & $10^{-14}$ & $10^{-2}$\\
  Air at normal pressure  & $10^{-31}$ & $10^{-19}$\\
  \noalign{\smallskip}\hline
\end{tabular}
\end{table}

\subsection{\label{sec:quant-brown-moti}Quantum Brownian motion}

A classic and extensively studied model of decoherence and dissipation is the one-dimensional motion of a particle weakly coupled to a thermal bath of noninteracting harmonic oscillators, a model known as \emph{quantum Brownian motion}. The self-Hamiltonian $H_E$ of the environment is given by
\begin{equation}
  \label{eq:sfsfjaa11}
  H_E = \sum_i \left( \frac{1}{2m_i}p_i^2 +
  \frac{1}{2}m_i\omega_i^2q_i^2 \right),  
\end{equation}
where $m_i$ and $\omega_i$ denote the mass and natural frequency of the $i$th oscillator, and $q_i$ and $p_i$ are the canonical position and momentum operators. The interaction Hamiltonian $H_\text{int}$ describes the bilinear coupling of the system's position coordinate $x$ to the positions $q_i$ of the environmental oscillators, $H_\text{int} = x \otimes \sum_i c_i q_i$, where the $c_i$ denote coupling strengths. This interaction represents the continuous environmental monitoring of the position coordinate of the system. 

The Born--Markov master equation describing the evolution of the density matrix $\rho_S(t)$ of the system is given by \cite{Hu:1992:om,Schlosshauer:2007:un}
\begin{align}
\label{eq:vjp32q22}
  \frac{\D}{\D t} \rho_S(t) 
  &= -\I \bigl[ H_S, \rho_S(t) \bigr] \notag\\ & \quad -
  \int_0^\infty \D \tau \, \bigl( \nu(\tau) \bigl[ x, \bigl[
      x(-\tau), \rho_S(t) \bigr]\bigr] 
    \notag\\ & \qquad \qquad  - \I \eta(\tau) \bigl[ x,
    \bigl\{ x(-\tau), \rho_S(t)
    \bigr\}\bigr] \bigr).
\end{align}
Here, $x(\tau)$ denotes the system's position operator in the interaction picture, $x(\tau) = \E^{\I H_S\tau} x\E^{-\I H_S\tau}$. The curly brackets $\{\,\cdot\,\, , \,\cdot\, \}$ in the second line denote the anticommutator $\{ A, B \} \equiv AB + BA$. The functions
\begin{align}
  \nu(\tau) &= \int_0^\infty \D \omega
  \, J(\omega) \coth
  \left(\frac{\omega}{2k T}\right) \cos \left(\omega\tau\right), \label{eq:vdjpoo17} \\
  \eta(\tau) &= \int_0^\infty \D \omega\, J(\omega)
  \sin\left(\omega\tau\right), \label{eq:ponol218}
\end{align}
are known as the \emph{noise kernel} and \emph{dissipation kernel}, respectively. The function $J(\omega)$, called the \emph{spectral density} of the environment, is given by 
\begin{equation}
\label{eq:vdfpmdmv16}
  J(\omega) \equiv \sum_i  \frac{c_i^2}{2m_i\omega_i} \delta(\omega-\omega_i).
\end{equation}
In general, spectral densities encapsulate the physical properties of the environment. One frequently replaces the collection of individual environmental oscillators by an (often phenomenologically motivated) continuous function $J(\omega)$ of the environmental frequencies $\omega$. 

If we specialize to the important case of the system represented by a harmonic oscillator with 
self-Hamiltonian
\begin{equation}
  \label{eq:sfsfsdfy7jaa11}
  H_S =  \frac{1}{2M}p^2 +
  \frac{1}{2}M\Omega^2x^2,
\end{equation}
the resulting Born--Markov master equation is
\begin{multline}
\label{eq:vfoinbnd9s27}
  \frac{\D}{\D t} \rho_S(t) 
  = -\I \bigl[ H_S + \frac{1}{2}M
    \widetilde{\Omega}^2 x^2, \rho_S(t) \bigr]
  - \I \gamma \bigl[ x, \bigr\{ p,
      \rho_S(t) \bigr\} \bigr] 
\\ - D \bigl[ x, \bigl[ x, \rho_S(t) \bigr]
\bigr] 
- f \bigl[ x, \bigl[ p, \rho_S(t) \bigr]
\bigr].
\end{multline}
The coefficients $\widetilde{\Omega}^2$, $\gamma$, $D$, and $f$ are defined as
\begin{subequations}\label{eq:jcsfr09355378}
\begin{align}
  \widetilde{\Omega}^2 &\equiv - \frac{2}{M} \int_0^\infty \D \tau \,
  \eta(\tau) \cos\left( \Omega \tau \right), \label{eq:caytcs1} \\
  \gamma &\equiv \frac{1}{M\Omega} \int_0^\infty \D \tau \,
  \eta(\tau) \sin\left( \Omega \tau \right), \label{eq:caytcs2} \\
  D &\equiv  \int_0^\infty \D \tau \,
  \nu(\tau) \cos\left( \Omega \tau \right), \label{eq:caytcs3}  \\
  f &\equiv - \frac{1}{M\Omega} \int_0^\infty \D \tau \,
  \nu(\tau) \sin\left( \Omega \tau \right). \label{eq:caytcs4} 
\end{align}
\end{subequations}
The first term on the right-hand side of Eq.~\eqref{eq:vfoinbnd9s27} represents the unitary dynamics of a harmonic oscillator whose natural frequency is shifted by $\widetilde{\Omega}$. The second term describes momentum damping (dissipation) at a rate proportional to $\gamma$, which depends only on the spectral density but not the temperature of the environment. The third term is of the Lindblad double-commutator form [see Eq.~\eqref{eq:lindbladc}] and describes decoherence of spatial coherences over a distance $\Delta X$ at a rate $D(\Delta X)^2$. Note that $D$ depends on both the spectral density $J(\omega)$ and the temperature $T$ of the environment. The fourth term also represents decoherence, but its influence on the dynamics of the system is usually negligible, especially at higher temperatures. In the long-time limit $\gamma t \gg 1$, the master equation \eqref{eq:vfoinbnd9s27} describes dispersion in position space given by
\begin{equation}
  \Delta X^2(t) = \frac{D}{2m^2 \gamma^2} t.
\end{equation}
That is, the width $\Delta X(t)$ of the ensemble in position space asymptotically scales as $\Delta X(t) \propto \sqrt{t}$, just as in classical Brownian motion; hence the term ``quantum Brownian motion.'' 

\begin{figure}
\centering
\vspace{.6cm}
\includegraphics[scale=0.7]{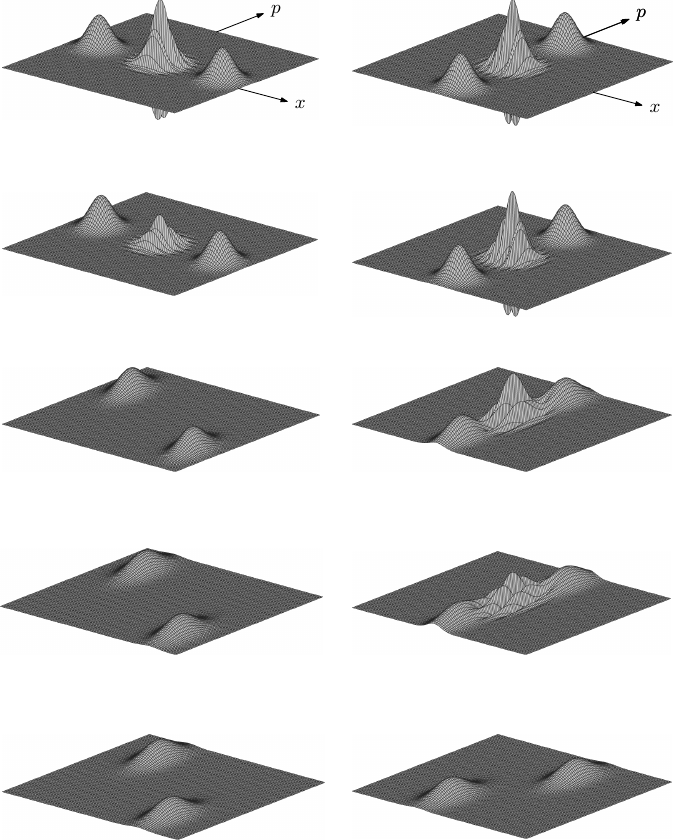}
\vspace{.6cm}
\caption{Evolution of superpositions of Gaussian wave
  packets in quantum Brownian motion as studied in Ref.~\cite{Paz:1993:ta}, visualized in the Wigner representation. Time increases from top to bottom. In the left column, the initial wave packets are separated in position; in the right column, the separation is in momentum. }
\label{fig:gaussmov}
\end{figure}

Figure~\ref{fig:gaussmov} shows the time evolution of position-space and momentum-space superpositions of two Gaussian wave
packets in the Wigner picture, as described by Eq.~\eqref{eq:vfoinbnd9s27} \cite{Paz:1993:ta}. Interference between the two wave packets is represented by oscillations between the direct peaks. The interaction with the environment damps these oscillations. The damping occurs on different timescales for the two initial conditions. While the momentum coordinate is not directly monitored by the environment, the intrinsic dynamics, through their creation of spatial superpositions from superpositions of momentum, result in decoherence in momentum space. This interplay of environmental monitoring and intrinsic dynamics leads to the emergence of pointer states that are minimum-uncertainty Gaussians (coherent states) well-localized in both position and momentum, thus approximating classical points in phase space \cite{Kubler:1973:ux,Paz:1993:ta,Zurek:1993:pu,Zurek:2002:ii,Diosi:2000:yn,Joos:2003:jh,Eisert:2003:ib}.

Let us consider the important case of an ohmic spectral density $J(\omega) \propto \omega$ with a high-frequency cutoff $\Lambda$,
\begin{equation}
  \label{eq:pojsvsddsjldfv1}
  J(\omega) = \frac{2M\gamma_0}{\pi} \omega
  \frac{\Lambda^2}{\Lambda^2 + \omega^2}.
\end{equation}
In the limit of a high-temperature environment ($kT \gg \Omega$ and $kT \gg \Lambda$), we arrive at the \emph{Caldeira--Leggett master equation} \cite{Caldeira:1983:on},
\begin{multline}
\label{eq:vfnbcclclnd9s27}
  \frac{\D}{\D t} \rho_S(t) 
  = -\I \bigl[ H'_S, \rho_S(t) \bigr]
  - \I \gamma_0 \bigl[ x, \bigl\{ p,
      \rho_S(t) \bigr\} \bigr] 
\\ - 2 M\gamma_0 k T \bigl[ x, \bigl[ x, \rho_S(t) \bigr]
\bigr],
\end{multline}
where 
\begin{equation}
  H'_S = H_S + \frac{1}{2}M
  \widetilde{\Omega}^2 x^2 = \frac{1}{2M}p^2 +
  \frac{1}{2}M\left[ \Omega^2 - 2\gamma_0 \Lambda \right]x^2
\end{equation}
is the frequency-shifted Hamiltonian $H'_S$ of the system. This equation has been widely and successfully used to model decoherence and dissipation processes, even in cases where the assumptions were not strictly fulfilled (for example, in quantum-optical settings, where often $kT \lesssim \Lambda$ \cite{Walls:1985:lm}).

In the position representation, the final term on the right-hand side of Eq.~\eqref{eq:vfnbcclclnd9s27} can be written as
\begin{equation}
  \label{eq:fsdojgdj1}
  -  \gamma_0 \left( \frac{x-x'}{\lambda_\text{dB}} \right)^2 \rho_S(x,x',t),
\end{equation}
where $\lambda_\text{dB} = (2 M k T)^{-1/2}$ is the thermal de Broglie wavelength. This term describes spatial localization with a decoherence rate $\tau_{\abs{x-x'}}^{-1}$ given by \cite{Zurek:1986:uz}
\begin{equation}
\label{eq:odijsvuhfsw21}
  \tau_{\abs{x-x'}}^{-1} = \gamma_0 \left(
    \frac{x-x'}{\lambda_\text{dB}} \right)^2.
\end{equation}
This is Eq.~\eqref{eq:daf12}, and as discussed there, given that $\lambda_\text{dB}$ is extremely small for macroscopic and even mesoscopic objects, we see that superpositions of macroscopically separated center-of-mass positions will typically be decohered on timescales many orders of magnitude shorter than the dissipation (relaxation) timescale $\gamma^{-1}_0$. Over timescales on the order of the decoherence time, we may therefore often neglect the dissipative term in Eq.~\eqref{eq:vfnbcclclnd9s27}, leading to the pure-decoherence master equation
\begin{equation}
\label{eq:vfnbcclasclnd9s27}
  \frac{\D}{\D t} \rho_S(t)  = -\I \bigl[
  H'_S, \rho_S(t) \bigr] 
 - 2 M\gamma_0 k T \bigl[ x, \bigl[ x, \rho_S(t) \bigr]
\bigr].
\end{equation}

\subsection{\label{sec:spin-boson-models}Spin--boson models}

In the spin--boson model, a qubit interacts with an environment of harmonic oscillators. The seminal review paper by Leggett et al.\ \cite{Leggett:1987:pm} discusses the dynamics of the spin--boson model in great detail. 

Let us first consider a simplified spin--boson model where the  self-Hamiltonian of the system is taken to be  $H_S = \frac{1}{2} \omega_0 \sigma_z$, with eigenstates $\ket{0}$ and $\ket{1}$. In contrast with the more general case discussed below, this Hamiltonian does not include a tunneling term $-\frac{1}{2} \Delta_0 \sigma_x$, and thus $H_S$ does not generate any nontrivial intrinsic dynamics. We employ the familiar self-Hamiltonian, Eq.~\eqref{eq:sfsfjaa11}, for an environment of harmonic oscillators, and choose the bilinear interaction Hamiltonian $H_\text{int} =   \sigma_z \otimes \sum_i c_i q_i$. Using the raising and lowering operators $a^\dagger$ and $a$, we can recast the total Hamiltonian as
\begin{equation}\label{eq:h-ssb}
  H =  \frac{1}{2} \omega_0 \sigma_z 
  + \sum_i \omega_i a_i^\dagger a_i + \sigma_z \otimes  \sum_i 
  \left( g_ia_i^\dagger + g_i^* a_i \right).
\end{equation}
Note that since $\bigl[ H, \sigma_z \bigr] = 0$, no transitions between $\ket{0}$ and $\ket{1}$ can be induced by $H$. There is no energy exchange between the system and the environment, and we therefore deal with a model of decoherence without dissipation. Such a model is a good representation of rapid decoherence processes during which the amount of dissipation is negligible, as is often the case in physical applications. The resulting evolution can be solved exactly \cite{Schlosshauer:2007:un}. For an ohmic spectral density with a high-frequency cutoff, it is found that superpositions of the form $\alpha\ket{0}+\beta\ket{1}$ are exponentially decohered on a timescale set by the thermal correlation time $(kT)^{-1}$ of the environment.

Inclusion of a tunneling term $-\frac{1}{2} \Delta_0 \sigma_x$ yields the general spin--boson model defined by the
Hamiltonian
\begin{multline}\label{eq:h-sbhdcskgsf}
H = \frac{1}{2} \omega_0 \sigma_z - 
\frac{1}{2} \Delta_0 \sigma_x 
+  \sum_i \left( \frac{1}{2m_i} p_i^2 + \frac{1}{2} m_i \omega_i^2 q_i^2 
   \right) \\+ \sigma_z \otimes \sum_i
  c_i q_i.
\end{multline}
The rich non-Markovian dynamics of this model have been analyzed in Refs.~\cite{Leggett:1987:pm,Weiss:1999:tv}. The particular dynamics strongly depend on the various parameters, such as the temperature of the environment, the form of the spectral density (subohmic, ohmic, or supraohmic), and the system--environment coupling strength. For each parameter regime, a characteristic dynamical behavior emerges: localization, exponential or incoherent relaxation, exponential decay, and strongly or weakly damped coherent oscillations \cite{Leggett:1987:pm}.

In the weak-coupling limit, one can derive the Born--Markov master equation in much the same way as in the case of quantum Brownian motion (note the similar structure of the Hamiltonians). The result is (see Ref.~\cite{Schlosshauer:2007:un} for details)
\begin{multline}
\label{eq:vjp32gbntrkh22}
\frac{\D}{\D t} \rho_S(t) = -\I \left(
  H'_S \rho_S(t) - {\rho}_S(t)
  H'^\dagger_S \right) \\ - \widetilde{D} \left[
  \sigma_z, \left[ \sigma_z, \rho_S(t)
  \right]\right]  + \zeta \sigma_z
\rho_S(t)\sigma_y + \zeta^* \sigma_y
\rho_S(t)\sigma_z.
\end{multline}
The first term on the right-hand side of the master equation \eqref{eq:vjp32gbntrkh22} represents the evolution under the environment-shifted self-Hamiltonian $H'_S$, the second term corresponds to decoherence in the $\sigma_z$ eigenbasis of the system at a rate given by $\widetilde{D}$, and the last two terms describe the decay of the two-level system. $H'_S$ is the renormalized (and in general non-Hermitian) Hamiltonian of the
system. The coefficients $\zeta^*$, $\widetilde{D}$, $\widetilde{f}$, and $\widetilde{\gamma}$ are given by
\begin{subequations}
\begin{align}
  \zeta^* &= \widetilde{f} - \I \widetilde{\gamma}, \\
  \widetilde{D} &= \int_0^\infty \D \tau \,
  \nu(\tau) \cos\left( \Delta_0 \tau \right), \\
  \widetilde{f} &=  \int_0^\infty \D \tau \, \nu(\tau)
  \sin\left( \Delta_0 \tau
  \right), \label{eq:dfvsih882e46b}\\
  \widetilde{\gamma} &=  \int_0^\infty \D \tau \, \eta(\tau)
  \sin\left( \Delta_0 \tau \right),\label{eq:dfvsih882e46c}
\end{align}
\end{subequations}
with the noise and the dissipation kernels $\nu(\tau)$ and $\eta(\tau)$ taking the same form as in quantum Brownian motion [see Eqs.~\eqref{eq:vdjpoo17} and \eqref{eq:ponol218}]. 

\subsection{\label{sec:spin-envir-models}Spin-environment models}

A qubit linearly coupled to a collection of other qubits---known also as a \emph{spin--spin model}---is often a good model of a single two-level system, such as a superconducting qubit, strongly coupled to a low-temperature environment \cite{Prokofev:2000:zz,Dube:2001:zz}. The model of a harmonic oscillator interacting with a spin environment may be relevant to the description of decoherence and dissipation in quantum-nanomechanical systems and micron-scale ion traps \cite{Schlosshauer:2008:os}. For details on the theory of spin-environment models, see Refs.~\cite{Dube:2001:zz,Stamp:1998:im,Prokofev:1995:ab,Prokofev:1993:aa}.

A simple version of a spin--spin model is described by the total Hamiltonian
\begin{align}\label{eq:jhdsgwuygfurwb}
  H = H_S + H_\text{int}& = - \frac{1}{2}
  \Delta_0 \sigma_x + \frac{1}{2} \sigma_z \otimes
  \sum_{i=1}^N g_i
  \sigma_z^{(i)} \notag \\ &
  \equiv - \frac{1}{2} \Delta_0 \sigma_x + \frac{1}{2}
  \sigma_z \otimes E.
\end{align}
Here, $H_S$ represents the intrinsic dynamics given by a tunneling term, while $H_\text{int}$ describes the environmental monitoring of the observable $\sigma_z$.

The model can be solved exactly \cite{Dobrovitski:2003:az,Cucchietti:2005:om}, and the resulting dynamics illustrate the dependence of the preferred basis on the relative strengths of the self-Hamiltonian of the system and the interaction Hamiltonian. The preferred basis emerges as the local basis that is most robust under the \emph{total} Hamiltonian. When the interaction Hamiltonian dominates over the self-Hamiltonian, the pointer states are found to be eigenstates of the interaction Hamiltonian, in agreement with the commutativity criterion, Eq.~\eqref{eq:dhvvsdnbbfvs27}. Conversely, when the modes of the environment are slow and the self-Hamiltonian dominates the evolution of the system (the quantum limit of decoherence \cite{Paz:1999:vv}), the pointer states are the eigenstates of the Hamiltonian of the system. 

In the weak-coupling limit, spin environments can be mapped onto oscillator environments \cite{Feynman:1963:jj,Caldeira:1993:bz}. Specifically, the reduced dynamics of a system weakly coupled to a spin environment can be described by the system coupled to an \emph{equivalent} oscillator environment described by an explicitly temperature-dependent spectral density of the form
\begin{equation}
\label{eq:vslkfvfgyiJA2}
J_\text{eff}(\omega, T) \equiv J(\omega)
\tanh\left(\frac{\omega}{2kT}\right),
\end{equation}
where $J(\omega)$ is the original spectral density of the spin environment. (See Sec.~5.4.2 of Ref.~\cite{Schlosshauer:2007:un} for details and examples.)

\section{\label{sec:decoh-errcorr}Qubit decoherence, quantum error correction, and error avoidance}

Quantum computation and quantum information processing rely on coherent superpositions of mesoscopically or macroscopically distinct states that are highly susceptible to decoherence. Avoiding, controlling, and mitigating decoherence is therefore of paramount importance. While the qubits need to be protected from detrimental environmental interactions, we also need to be able to control and measure them via a macroscopic apparatus. The formidable challenge of designing a quantum computer consists of meeting both demands in a balanced way. Even so, decoherence induced by interactions with the environment and the control apparatus, as well as noise due to faulty gate operations, will likely be too strong to allow for useful quantum computations to be carried out \cite{Miquel:1997:zz,Miquel:1996:ra}. What is also needed is an active mitigation of the effects of decoherence through active quantum error correction \cite{Steane:1996:cd,Shor:1995:rx,Steane:2001:dx,Knill:2002:rx,Nielsen:2000:tt}.

We may distinguish two limiting cases for modeling decoherence in qubits. The first limit is that of \emph{independent qubit decoherence}. Here, each qubit couples independently to its own environment, without any interactions between these environments. For example, this may be the case if the qubits are spatially well-separated (relative to the typical coherence length of the environment) and only couple to their immediate surroundings. Then the error processes affecting the qubits will be completely uncorrelated. Thus, if the probability of a particular error to affect one qubit is $p$, the probability of this error to occur in $K$ qubits will be $p^K$. Many error-correcting schemes are only efficient in correcting such single-qubit errors, and thus the assumption of independent decoherence frequently underlies these schemes. This assumption, however, is unrealistic when the qubits are located spatially close to each other. In this case, all qubits approximately feel the same environment, and it is likely that errors will become correlated among multiple qubits. The limiting case corresponding to this situation is that of \emph{collective qubit decoherence}, in which all qubits couple to exactly the same environment.

\subsection{\label{sec:corr-decoh-induc}Correction of decoherence-induced quantum errors}

Consider a single qubit $S$, initially described by a pure state $\ket{\psi}$ and interacting with an environment $E$. One can show that an arbitrary evolution of the combined qubit--environment state can always be written in the form
\begin{equation} \label{eq:qcerrc} \ket{\psi}
  \ket{e_0}  \, \longrightarrow \, I
  \ket{\psi} \ket{e_I} + \sum_{s= x,y,z }
  \left( \sigma_s \ket{\psi} \right)
  \ket{e_s},
\end{equation}
where the Pauli operators $\sigma_s$ act on the Hilbert space of $S$, and $\ket{e_I}$ and $\{ \ket{e_s} \}$ are environmental states that are not necessarily orthogonal or normalized. Thus, any influence of the environment on the qubit can be expressed simply in terms of a weighted sum of the Pauli operators and the identity operator acting on the original state of the qubit. The effects of $\sigma_x$ and $\sigma_z$ on the qubit state are often referred as a \emph{bit-flip error} and \emph{phase-flip error}, respectively. If we restrict our attention to environmental entanglement and the resulting decoherence effects, then only phase-flip errors need to be taken into account. 

For $N$ qubits, Eq.~\eqref{eq:qcerrc} generalizes to
\begin{equation}\label{eq:qcerrcN}
  \ket{\psi} \ket{e_0} \, \longrightarrow\,
  \sum_{i} 
  \left( E_i \ket{\psi} \right) \ket{e_i} .
\end{equation}
Here $\ket{\psi}$ is the initial $N$-qubit state, and the \emph{error operators} $E_i$ are tensor products of $N$ operators involving identity and Pauli
operators. Equation \eqref{eq:qcerrcN} represents a worst-case scenario. In many cases, simplified versions can be used. One important case is that of \emph{partial decoherence}. Here, only a small number $K < N$ of qubits become entangled with the environment between two successive applications of an error-correcting mechanism. Then it will be sufficient to restrict our attention to the $2^K$ possible error operators made up of at most $K$ operators $\sigma_z$ and $N-K$ identity operators. In the case of independent qubit decoherence, we only need to consider a collection of independent phase-flip errors acting on single qubits, represented by error operators of the form $E = I\otimes \cdots \otimes I\otimes \sigma_z \otimes I\otimes \cdots \otimes I$. 

Given the entangled state on the right-hand side of Eq.~\eqref{eq:qcerrcN}, the goal of quantum error correction is to restore the initial (unknown) state $\ket{\psi}$. We let an ancilla, described by an initial state $\ket{a_0}$, interact with the qubit system such that
\begin{equation}\label{eq:errfsyn}
  \ket{a_0} \left[ \sum_{i} \left( E_i   \ket{\psi} \right)
    \ket{e_i} \right] \, \longrightarrow \, 
  \sum_{i} \ket{a_i} \left( E_i \ket{\psi} \right)
  \ket{e_i}.
\end{equation}
Let us assume that the ancilla states $\ket{a_i}$ are at least approximately mutually orthogonal, such that they can be distinguished by measurement. We now measure the observable $O_A = \sum_i a_i \ketbra{a_i}{a_i}$ on the ancilla, with $a_i \not= a_j$ for $i\not= j$. The projective measurement will yield a particular outcome, say, $a_k$, and lead to the reduction of the entangled state,
\begin{equation}
\label{eq:fijvdnjvcsh411}
\sum_{i} \ket{a_i} \left( E_i \ket{\psi} \right)
\ket{e_i} \, \longrightarrow \, \ket{a_k}
\left( E_k \ket{\psi} \right) 
\ket{e_k}.
\end{equation}
The outcome $a_k$ of the measurement tells us the countertransformation needed to restore the initial qubit state. Applying $E_k^{-1}=E_k^\dagger$ to the system gives
\begin{equation}
  \ket{a_k} \left( E_k \ket{\psi} \right) \ket{e_k} \,
  \xrightarrow{E_k^{-1}} \,  
  \ket{a_k} \ket{\psi} \ket{e_k}.
\end{equation}
Note that, as required in order to avoid introducing additional decoherence in the computational basis of the qubit system, we have obtained no information whatsoever about the state of the system. 

This account of quantum error correction has been highly idealized. Let us mention three complications. First, it is impossible to design an interaction between the computational qubits and the ancilla that would allow us to distinguish, by measuring the ancilla, between \emph{all} possible errors. Second, in realistic settings the error operators $E_i$ may be very complex, and it remains to be seen whether and how the corresponding countertransformations can be applied without introducing significant additional decoherence. Third, the ancilla qubits are physically similar to the computational qubits and can therefore be expected to be equally prone to environmental interactions (and thus decoherence) as the computational qubits themselves. Since the inclusion of ancilla qubits increases the total number of qubits in the quantum computer, and since decoherence rates typically scale exponentially with the size of
the system, it will require sophisticated experimental designs to ensure not only that quantum error correction works in practice, but also that it does not aggravate the problem of qubit decoherence.

\subsection{Quantum computation on decoherence-free subspaces}

We introduced the concept of decoherence-free subspaces (DFS) \cite{Palma:1996:yy,Lidar:1998:uu,Zanardi:1997:yy,Zanardi:1997:tv,Zanardi:1998:oo,Lidar:1999:fa,Bacon:2000:yy,Duan:1998:yb,Zanardi:2001:oo,Knill:2000:aa}, or
pointer subspaces \cite{Zurek:1982:tv}, in Sec.~\ref{sec:dfs}. DFS allow us to encode quantum information in ``quiet corners'' of the Hilbert space to protect it from environmental effects. In contrast with quantum error correction, DFS prevent errors from happening in the first place and thus represent a strategy for intrinsic error avoidance.

The two limiting cases of independent qubit decoherence and collective qubit decoherence delineate the limits on the size of a DFS.  To illustrate this relationship, let us consider the case of collective decoherence of an $N$-qubit system interacting with an oscillator bath \cite{Palma:1996:yy,Duan:1998:yb,Reina:2002:ta,Zanardi:1997:yy,Zanardi:1998:oo}. The interaction Hamiltonian for this generalized spin--boson model is taken to be [compare Eq.~\eqref{eq:h-ssb}] 
\begin{equation}
  \label{eq:dadrestinpeacdh}
  H_\text{int} =  \sum_{i=1}^N  \sigma_z^{(i)} \otimes \sum_j
  \left( g_{ij}a_j^\dagger + g_{ij}^* a_j \right) \equiv
  \sum_{i=1}^N  \sigma_z^{(i)} \otimes E_i.
\end{equation}
The assumption of collective decoherence implies that the couplings $g_{ij}$ (and thus the environment operators $E_i$) must be independent of the index $i$. Then Eq.~\eqref{eq:dadrestinpeacdh} becomes
\begin{equation}
\label{eq:dadrestinpeacfxndh}
H_\text{int} =  \left( \sum_{i} \sigma_z^{(i)} \right)
\otimes E \equiv
S_z \otimes E.
\end{equation}
Recall that a DFS is spanned by a degenerate set of eigenstates of the system operators $S_\alpha$ of the interaction Hamiltonian [see Eq.~\eqref{eq:OIbvsrhjkbvsfljvh9}]. Thus, in our case the DFS will be spanned by degenerate eigenstates of the collective spin operator $S_z$. Any $N$-qubit product state of the computational basis states $\ket{0}$ and $\ket{1}$ (the eigenstates of $\sigma_z$ with eigenvalues $+1$ and $-1$, respectively) will be an eigenstate of $S_z$. There are $2N+1$ different possible integer eigenvalues $m$, ranging from $m=-N$ (corresponding to the basis state $\ket{1\cdots 1}$) to $m=+N$ (corresponding to $\ket{0\cdots 0}$). The largest number of mutually orthogonal computational-basis states with the same eigenvalue $m$ of $S_z$ is given by the set $\mathfrak{S}_0$ of basis states with $m=0$, i.e., those with $N/2$ qubits in the state $\ket{0}$. There are $n_0 = \binom{N}{N/2}$ such states in this set, spanning a DFS of dimension $n_0$. For large values of $N$, we can approximate the binomial coefficient using Stirling's formula,
\begin{equation}
 \log_2 \binom{N}{N/2} \approx N - \frac{1}{2} \log_2 (\pi N/2) \,\,
 \xrightarrow{N \gg 1} \,\, N.
\end{equation}
Therefore, in the limiting case of collective decoherence, the dimension of our DFS approaches the dimension of the original
Hilbert space, and the encoding efficiency approaches unity. For example, for $N=4$ qubits, the set 
\begin{equation}
  \label{eq:39}
  \mathfrak{S}_0 = \left\{\, \ket{0011}, \ket{0101}, \ket{0110}, \ket{1001}, \ket{1010}, 
\ket{1100} \, \right\}
\end{equation}
spans a maximum-size DFS of dimension six, to be compared with the dimension of the original Hilbert space, which is $2^4 = 16$. Thus, given the model for
collective decoherence considered here, using four physical qubits we can encode up to two logical qubits in a DFS (since encoding three logical qubits would already require a DFS of dimension $2^3=8$). 

As mentioned in Sec.~\ref{sec:dfs}, the existence of a DFS corresponds to a dynamical symmetry. Our model represents a case of perfect dynamical symmetry, since the system--environment interaction, Eq.~\eqref{eq:dadrestinpeacfxndh}, is completely symmetric with respect to any permutations of the qubits, thereby leading to a DFS of maximum size. What happens if the symmetry is broken by additional small independent coupling terms? It has been shown \cite{Lidar:1998:uu,Bacon:1999:aq} that, to first order in the perturbation strength, the \emph{storage} of quantum information in DFS is stable to such perturbations to all orders in time, but that the \emph{processing} of such quantum information encoded in DFS is robust only to first order in time. 

In the case of purely independent qubit decoherence, the environment operators $E_i$ appearing in Eq.~\eqref{eq:dadrestinpeacdh} will now differ from one another. To find a DFS, we follow the usual strategy [see Eq.~\eqref{eq:OIbvsrhjkbvsfljvh9}] of determining a set of orthonormal basis states $\{ \ket{s_i} \}$ such that
\begin{multline}
  \label{eq:16}
  \left[ I^{(1)} \otimes \cdots \otimes I^{(j-1)} \otimes
    \sigma_z^{(j)} \otimes I^{(j+1)} \otimes \cdots
    \otimes I^{(N)} \right] \ket{s_i}\\ = \lambda^{(j)} \ket{s_i} 
\end{multline}
for all $i$ and $1 \le j \le N$. The only state fulfilling this eigenvalue problem is $\ket{0\cdots 0}$. Since we need at least a two-dimensional subspace to encode a single logical qubit, the case of independent decoherence in the spin--boson model does not allow for the existence of a DFS for quantum computation. In the language of pointer subspaces, there is only a single exact pointer state, and this environment-superselected preferred state of the system will be the ground state $\ket{0\cdots 0}$.

In realistic settings, neither the assumption of purely independent decoherence nor the limit of entirely collective decoherence will be entirely appropriate. We can, however, use a DFS to protect the qubits from collective decoherence effects, and we can recover from single-qubit errors due to independent decoherence using active error-correction methods. These two approaches can be \emph{concatenated} \cite{Lidar:1999:fa} to enable universal fault-tolerant quantum computation even when the restriction to single-qubit errors is dropped \cite{Bacon:2000:yy,Lidar:2001:oo}.

\subsection{Environment engineering and dynamical decoupling}

For reasonably large DFS to exist, the system--environment interaction must exhibit a sufficiently high degree of symmetry. Such symmetries are unlikely to arise naturally in typical experimental settings.

One way of overcoming this limitation is based on \emph{environment engineering}. Here, one tries to generate
certain symmetries in the structure of the system--environment interactions. For example, an appropriately engineered symmetrization could make superposition states in Bose--Einstein condensates correspond to (approximate) degenerate eigenstates of the interaction Hamiltonian, in which case such states would lie within a DFS, thereby significantly enhancing their longevity \cite{Dalvit:2000:bb}. In ion traps, changing the parameters in the effective interaction Hamiltonian for the trapped ion allows one to select different pointer subspaces and thereby control into which DFS the trapped ion is driven \cite{Poyatos:1996:um,Myatt:2000:yy,Turchette:2000:aa,Carvalho:2001:ua}. 

Another approach to the active creation of DFS is known as \emph{dynamical decoupling} \cite{Viola:1998:uu,Viola:1999:zp,Zanardi:1999:oo,%
  Viola:2000:pp,Wu:2002:aa,Wu:2002:bb}. Here time-dependent modifications are introduced into the Hamiltonian of the system that counteract the influence of the environment. These modifications take the form of sequences of rapid projective measurements or strong control-field pulses acting on the system (``quantum bang-bang control'' \cite{Viola:1998:uu}). Even if the structure of the system--environment interaction Hamiltonian is not known, decoherence can be suppressed arbitrarily well in the limit of an infinitely fast rate of the decoupling control field, thus dynamically creating a DFS (which then represents a dynamically decoupled subspace).  In the realistic case of a finite control rate, sufficient (albeit imperfect) protection from decoherence can be achieved via this decoupling technique, provided the control rate is larger than the fastest timescale set by the rate of formation of environmental entanglement.

\section{\label{sec:exper-observ-decoh}Experimental studies of decoherence}

Decoherence, of course, happens all around us, and in this sense its consequences are readily observed. But what we would like to do is to be able to experimentally study the \emph{gradual} and \emph{controlled} action of decoherence. In this endeavor, several obstacles have to be overcome. We need to prepare the system in a superposition of mesoscopically or even macroscopically distinguishable states with a sufficiently long decoherence time such that the gradual action of decoherence can be resolved. We must be able to monitor decoherence without introducing a significant amount of additional, unwanted decoherence. We would also like to have sufficient control over the environment so we can tune the strength and form of its interaction with the system. Starting in the mid-1990s, several such experiments have been performed, for example, using cavity QED \cite{Raimond:2001:aa}, mesoscopic molecules \cite{Arndt:2005:mi}, and superconducting systems such as SQUIDs and Cooper-pair boxes \cite{Leggett:2002:uy}. Bose--Einstein condensates \cite{Kaiser:2001:tm} and quantum nanomechanical systems \cite{Blencowe:2004:mm,Aspelmeyer:2013:aa} are promising candidates for future experimental tests of decoherence.

These experiments are important for several reasons. They are impressive demonstrations of the possibility of generating nonclassical quantum states in mesoscopic and macroscopic systems. They show that the quantum--classical boundary is smooth and can be shifted by varying the relevant experimental parameters. They allow us to test and improve decoherence models, and they help us design devices for quantum information processing that are good at evading the detrimental influence of the environment. Finally, such experiments may be used to test quantum mechanics itself \cite{Leggett:2002:uy}. Such tests require sufficient shielding of the system from decoherence so that an observed (full or partial) collapse of the wavefunction could be unambigously attributed to some novel nonunitary mechanism in nature, such as those proposed in dynamical reduction models \cite{Bassi:2003:yb,Adler:2007:um,Bassi:2010:aa}. This shielding, however, is difficult to implement in practice, because the large number of particles required for the reduction mechanism to become effective will also lead to strong decoherence \cite{Tegmark:1993:uz,Nimmrichter:2013:aa}. The superpositions realized in current experiments are still not sufficiently macroscopic to rule out collapse theories, although it has been demonstrated \cite{Nimmrichter:2011:pr} that matter-wave interferometry with large molecular clusters (in the mass range between $10^6$ and $\unit[10^8]{amu}$) would be able to test the collapse theories proposed in Refs.~\cite{Adler:2007:um,Bassi:2010:aa}; such experiments may soon become technologically feasible \cite{Hornberger:2012:ii}. 

\subsection{\label{sec:atoms-cavity}Atoms in a cavity}

In 1996 Brune et al.\ generated a superposition of radiation fields with classically distinguishable phases involving several photons \cite{Brune:1996:om,Raimond:2001:aa,Kaiser:2001:tm}. This experiment was the first to realize a mesoscopic Schr\"odinger-cat state and allowed for the controlled observation and manipulation of its decoherence. A rubidium atom is prepared in a superposition of energy eigenstates $\ket{g}$ and $\ket{e}$ corresponding to two circular Rydberg states. The atom enters a cavity $C$ containing a radiation field containing a few photons. If the atom is in the state $\ket{g}$, the field remains unchanged, whereas if it is in the state  $\ket{e}$, the coherent state $\ket{\alpha}$ of the field undergoes a phase shift $\phi$, $\ket{\alpha} \longrightarrow \ket{\E^{i\phi} \alpha}$; the experiment achieved $\phi \approx \pi$. An initial superposition of the atom is therefore amplified into an entangled atom--field state of the form $\frac{1}{\sqrt{2}} \left( \ket{g} \ket{\alpha} + \ket{e} \ket{-\alpha} \right)$. The atom then passes through an additional cavity, further transforming the superposition. Finally, the energy state of the atom is measured. This disentangles the atom and the field and leaves the latter in a superposition of the mesoscopically distinct states $\ket{\alpha}$ and $\ket{-\alpha}$.

To monitor the decoherence of this superposition, a second rubidium atom is sent through the apparatus. After interacting with the field superposition state in the cavity $C$, the atom will always be found in the same energy state as the first atom if the superposition has not been decohered. This correlation rapidly decays with increasing decoherence. Thus, by recording the measurement correlation as a function of the wait time $\tau$ between sending the first and second atom through the apparatus, the decoherence of the field state can be monitored. Experimental results were in excellent agreement with theoretical predictions \cite{Davidovich:1996:sa,Maitre:1997:tv}. It was found that decoherence became faster as the phase shift $\phi$ and the mean number $\bar{n}=\abs{\alpha}^2$ of photons in the cavity $C$ was increased. Both results are expected, since an increase in $\phi$ and $\bar{n}$ means that the components in the superposition become more distinguishable. Recent experiments have realized superposition states involving several tens of photons \cite{Auffeves:2003:za} and have monitored the gradual decoherence of such states \cite{Deleglise:2008:oo}.

\subsection{\label{sec:matt-wave-interf}Matter-wave interferometry}

\begin{figure}[t]
\centering
\vspace{.cm}
\includegraphics[scale=0.85]{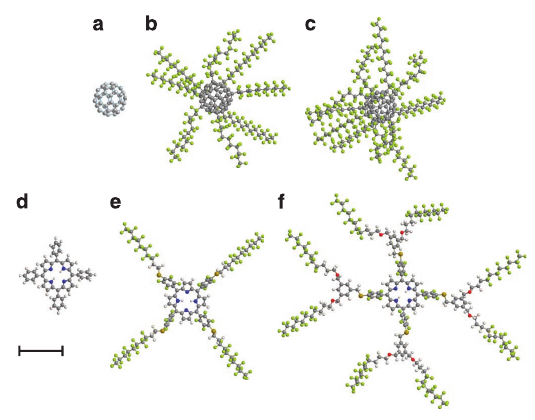}\qquad\includegraphics[scale=1]{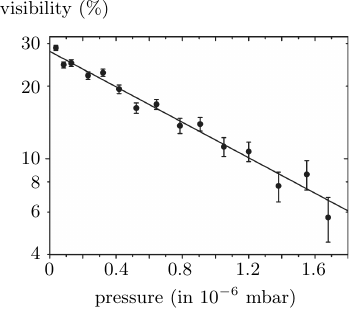}
\caption{\emph{Left:} Molecular clusters used in recent interference experiments, drawn to scale (the scale bar represents \unit[10]{\AA}). Figure from Ref.~\cite{Gerlich:2011:aa}. \emph{(a)} Fullerene C$_{60}$ ($m = \unit[720]{amu}$, 60 atoms). \emph{(b)}
Perfluoroalkylated nanosphere PFNS8 ($m = \unit[5672]{amu}$, 356 atoms). \emph{(c)} PFNS10 ($m = \unit[6910]{amu}$, 430 atoms). \emph{(d)} Tetraphenylporphyrin TPP ($m = \unit[614]{amu}$, 78 atoms). \emph{(e)} TPPF84 ($m = \unit[2814]{amu}$, 202 atoms). \emph{(f)} TPPF152 ($m = \unit[5310]{amu}$, 430 atoms). \emph{Right:} Visibility of interference fringes of C$_{70}$ fullerenes as a function of the pressure of the background gas. Measured values (circles) agree well with the theoretical prediction (solid line) \cite{Hornberger:2003:un,Hornberger:2004:bb,Hornberger:2005:mo} describing an exponential decay of visibility with pressure. Figure adapted from
  Ref.~\cite{Hackermuller:2003:uu}.}
\label{fig:c70-vis}
\end{figure}

In these experiments (see Ref.~\cite{Hornberger:2012:ii} for a review), spatial interference patterns are demonstrated for mesoscopic molecules ranging from fullerenes \cite{Arndt:1999:rc} to molecular clusters involving hundreds of atoms, with a total size of up to \unit[60]{\AA} and masses of several thousand amu (see Fig.~\ref{fig:c70-vis}) \cite{Gerlich:2011:aa,Eibenberger:2013:az}. Since the de~Broglie wavelength of such molecules is on the order of picometers, standard double-slit interferometry is out of reach. Instead, the experiments make use of the Talbot effect, an interference phenomenon in which a plane wave incident on a diffraction grating creates an image of the grating at multiples of a distance $L$ behind the grating. In the experiment, the molecular density (at a macroscopic distance $L$ from the grating) is scanned along the direction perpendicular to the molecular beam. An oscillatory density pattern (corresponding to the image of the slits in the grating) is observed, confirming the existence of coherence and interference between the different paths of each individual molecule passing through the grating. Recent experiments have used an improved version of the original Talbot--Lau setup \cite{Gerlich:2007:om}, as well as optical ionization gratings \cite{Haslinger:2013:ii}.

Decoherence is measured as a decrease of the visibility of this pattern (Fig.~\ref{fig:c70-vis}). The controlled decoherence due to collisions with background gas particles \cite{Hackermuller:2003:uu,Hornberger:2003:tv} and due to emission of thermal radiation from heated molecules \cite{Hackermuller:2004:rd} has been observed, showing a smooth decay of visibility in agreement with theoretical predictions \cite{Hornberger:2003:un,Hornberger:2004:bb,Hornberger:2005:mo}. These successes have led to speculations that one could perform similar experiments using even larger particles such as proteins and viruses \cite{Hackermuller:2003:uu,Arndt:2002:bo} or carbonaceous aerosols \cite{Hornberger:2006:tx}. Such experiments will be limited by collisional and thermal decoherence and by noise due to inertial forces and vibrations \cite{Hackermuller:2003:uu,Arndt:2002:bo,Hornberger:2006:tx}.

\subsection{\label{sec:superc-syst}Superconducting systems}

\begin{figure}
\raggedright (\emph{a}) 

\begin{center}
\includegraphics[scale=.8]{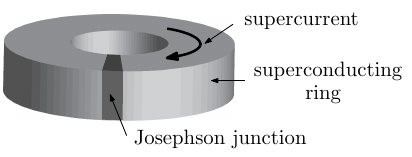}
\end{center}

\raggedright (\emph{b}) 

\begin{center}
\includegraphics[scale=.85]{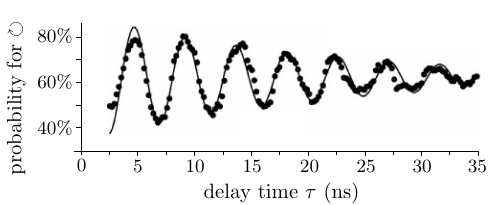}
\end{center}

\caption{\label{fig:squid}(\emph{a}) Schematic illustration of a SQUID. A superconducting ring is interrupted by Josephson junctions, leading to a dissipationless supercurrent. (\emph{b}) Decoherence of a superposition of clockwise and counterclockwise supercurrents in a superconducting qubit. The damping of the oscillation amplitude corresponds to the gradual loss of coherence from the system. Figure adapted from Ref.~\cite{Chiorescu:2003:ta}.}
\end{figure}

Superconducting quantum interference devices (SQUIDs) and Cooper-pair boxes have important applications in quantum information processing. A SQUID consists of a ring of superconducting material interrupted by thin insulating barriers, the Josephson junctions (Fig.~\ref{fig:squid}a). At sufficiently low temperatures, electrons of opposite spin condense into bosonic Cooper pairs.  Quantum-mechanical tunneling of Cooper pairs through the junctions leads to the flow of a resistance-free supercurrent around the loop (Josephson effect), which creates a magnetic flux threading the loop. The collective center-of-mass motion of a macroscopic number ($\sim 10^9$) of Cooper pairs can then be represented by a wave function labeled by a single macroscopic variable, namely, the total trapped flux $\Phi$ through the loop.  The two possible directions of the supercurrent define a qubit with basis states $\{\ket{\circlearrowright}, \ket{\circlearrowleft} \}$. By adjusting an external magnetic field, the SQUID can be biased such that the two lowest-lying energy eigenstates $\ket{0}$ and $\ket{1}$ are equal-weight superpositions of the persistent-current states $\ket{\circlearrowright}$ and $\ket{\circlearrowleft}$. 

Such superposition states involving $\mu$A currents were first experimentally observed in 2000 using spectroscopic measurements \cite{Friedman:2000:rr,Wal:2000:om}. Their decoherence was subsequently measured using Ramsey interferometry \cite{Chiorescu:2003:ta}, as follows. Two consecutive microwave pulses are applied to the system. During the delay time $\tau$ between the pulses, the system evolves freely. After application of the second pulse, the system is left in a superposition of $\ket{\circlearrowright}$ and $\ket{\circlearrowleft}$, with the relative amplitudes exhibiting an oscillatory dependence on $\tau$. A series of measurements in the basis $\{ \ket{\circlearrowright}, \ket{\circlearrowleft} \}$ over a range of delay times $\tau$ then allows one to trace out an oscillation of the occupation probabilities for $\ket{\circlearrowright}$ and $\ket{\circlearrowleft}$ as a function of $\tau$ (Fig.~\ref{fig:squid}b). The envelope of the oscillation is damped as a consequence of decoherence acting on the system during the free evolution of duration $\tau$. From the decay of the envelope we can infer the decoherence timescale; the original experiment gave 20~ns \cite{Chiorescu:2003:ta}, while subsequent experiments have achieved decoherence times of several $\mu$s \cite{Bertet:2005:un}.

Superpositions states and their decoherence have also been observed in superconducting devices whose key variable is charge (or phase), instead of the flux variable used in SQUIDs. Such Cooper-pair boxes consist of a small superconducting island onto which Cooper pairs can tunnel from a reservoir through a Josephson junction. Two different charge states of the island, differing by at least one Cooper pair, define the basis states. Coherent oscillations between such charge states were first observed in 1999 \cite{Nakamura:1999:ub}. In 2002, Vion et al.\ \cite{Vion:2002:oo} reported thousands of coherent oscillations with a decoherence time of 0.5~$\mu$s. Similar results have been obtained for phase qubits \cite{Yu:2002:yb,Martinis:2002:qq}, demonstrating decoherence times of several $\mu$s.

\section{\label{sec:impl-found-quant}Decoherence and the foundations of quantum mechanics}

Can decoherence address foundational problems? If so, which ones, and how? Addressing these subtle questions is beyond the scope of this review; a few brief remarks must suffice here. (See Refs.~\cite{Bacciagaluppi:2003:yz,Schlosshauer:2003:tv,Schlosshauer:2006:rw,Schlosshauer:2007:un} for in-depth discussions.)  Decoherence, at its heart, is a technical result concerning the dynamics and measurement statistics of open quantum systems. From this view, decoherence merely addresses a \emph{consistency problem}, by explaining how and when the quantum probability distributions approach the classically expected distributions. Since decoherence follows directly from an application of the quantum formalism to interacting quantum systems, it is not tied to any particular interpretation of quantum mechanics, nor does it supply such an interpretation, nor does it amount to a theory that could make predictions beyond those of standard quantum mechanics. 

The predictively relevant part of decoherence theory relies on reduced density matrices, whose formalism and interpretation presume the collapse postulate and Born's rule. If we understand the ``quantum measurement problem'' as the question of how to reconcile the linear, deterministic evolution described by the Schr\"odinger equation with the occurrence of random measurement outcomes, then decoherence has not solved this problem \cite{Schlosshauer:2003:tv,Schlosshauer:2007:un}. Decoherence does, however, address an aspect sometimes associated with the quantum measurement problem, namely the preferred-basis problem (at least in the sense described in Sec.~\ref{sec:envir-induc-supers}). Further explorations of the role of the environment, such as in quantum Darwinism (see Sec.~\ref{sec:prol-inform-quant}), can help illuminate fundamental questions concerning information transfer and amplification in the quantum setting.

Decoherence has been used to identify internal concistency issues in interpretations of quantum mechanics, and the picture associated with the decoherence process has sometimes been seen as suggestive of particular interpretations of quantum mechanics \cite{Schlosshauer:2003:tv, Bacciagaluppi:2003:yz}. Indeed, historically decoherence theory arose in the context of Zeh's \cite{Zeh:1970:yt} independent formulation of an Everett-style interpretation (see Ref.~\cite{Camilleri:2009:aq} for a historical analysis). Ultimately, however, it seems that certain interpretations simply may be more \emph{in need} of decoherence than others for defining their structure; see Ref.~\cite{Wallace:2010:im} for the example of an Everett-style interpretation \cite{Everett:1957:rw}. At the end of the day, any interpretation that does not involve entities, claims, or structures in contradiction with the predictions of decoherence theory (which is to say, with the predictions of quantum mechanics) will arguably remain viable.



\begin{thebibliography}{100}
\providecommand{\url}[1]{{#1}}
\providecommand{\urlprefix}{URL }
\expandafter\ifx\csname urlstyle\endcsname\relax
  \providecommand{\doi}[1]{DOI \discretionary{}{}{}#1}\else
  \providecommand{\doi}{DOI \discretionary{}{}{}\begingroup
  \urlstyle{rm}\Url}\fi

\bibitem{Zeh:1970:yt}
H.D. Zeh, Found. Phys. \textbf{1}, 69 (1970)

\bibitem{Zurek:1981:dd}
W.H. Zurek, Phys. Rev. D \textbf{24}, 1516 (1981)

\bibitem{Zurek:1982:tv}
W.H. Zurek, Phys. Rev. D \textbf{26}, 1862 (1982).
\newblock \doi{10.1103/PhysRevD.26.1862}

\bibitem{Paz:2001:aa}
J.P. Paz, W.H. Zurek, in \emph{Coherent Atomic Matter Waves, Les Houches
  Session LXXII}, \emph{Les Houches Summer School Series}, vol.~72, ed. by
  R.~Kaiser, C.~Westbrook, F.~David (Springer, Berlin, 2001), \emph{Les Houches
  Summer School Series}, vol.~72, pp. 533--614

\bibitem{Zurek:2002:ii}
W.H. Zurek, Rev. Mod. Phys. \textbf{75}, 715 (2003).
\newblock \doi{10.1103/RevModPhys.75.715}

\bibitem{Schlosshauer:2003:tv}
M.~Schlosshauer, Rev. Mod. Phys. \textbf{76}, 1267 (2004)

\bibitem{Bacciagaluppi:2003:yz}
G.~Bacciagaluppi, in \emph{The Stanford Encyclopedia of Philosophy}, ed. by
  E.N. Zalta (2012).
\newblock Online at
  http://plato.stanford.edu/archives/win2012/entries/qm-decoherence

\bibitem{Joos:2003:jh}
E.~Joos, H.D. Zeh, C.~Kiefer, D.~Giulini, J.~Kupsch, I.O. Stamatescu,
  \emph{Decoherence and the Appearance of a Classical World in Quantum Theory},
  2nd edn. (Springer, New York, 2003)

\bibitem{Schlosshauer:2007:un}
M.~Schlosshauer, \emph{Decoherence and the Quantum-to-Classical Transition}
  (Springer, Berlin/Heidelberg, 2007)

\bibitem{Schlosshauer:2019:qd}
M.~Schlosshauer, Phys. Rep. \textbf{831}, 1 (2019).
\newblock \doi{10.1016/j.physrep.2019.10.001}

\bibitem{Hornberger:2012:ii}
K.~Hornberger, S.~Gerlich, S.~Nimmrichter, P.~Haslinger, M.~Arndt, Rev. Mod.
  Phys. \textbf{84}, 157 (2012)

\bibitem{Raimond:2001:aa}
J.M. Raimond, M.~Brune, S.~Haroche, Rev. Mod. Phys. \textbf{73}, 565 (2001)

\bibitem{Leggett:2002:uy}
A.J. Leggett, J. Phys.: Condens. Matter \textbf{14}, R415 (2002)

\bibitem{Joos:1999:po}
E.~Joos, in \emph{Decoherence: Theoretical, Experimental, and Conceptual
  Problems}, ed. by P.~Blanchard, D.~Giulini, E.~Joos, C.~Kiefer, I.O.
  Stamatescu (Springer, Berlin, 2000), pp. 1--17

\bibitem{Schlosshauer:2011:un}
M.~Schlosshauer, K.~Camilleri, AIP Conf. Proc. \textbf{1327}, 26 (2011)

\bibitem{Kubler:1973:ux}
O.~K{\"u}bler, H.D. Zeh, Ann. Phys. (N.Y.) \textbf{76}, 405 (1973)

\bibitem{Joos:1985:iu}
E.~Joos, H.D. Zeh, Z. Phys. B: Condens. Matter \textbf{59}, 223 (1985)

\bibitem{Zurek:1986:uz}
W.H. Zurek, in \emph{Frontiers of Nonequilibrium Statistical Mechanics}, ed. by
  G.T. Moore, M.O. Scully (Plenum Press, New York, 1986), pp. 145--149.
\newblock First published in 1984 as Los Alamos report LAUR 84-2750

\bibitem{Hornberger:2009:aq}
K.~Hornberger, in \emph{Entanglement and Decoherence: Foundations and Modern
  Trends}, \emph{Lecture Notes in Physics}, vol. 768, ed. by A.~Buchleitner,
  C.~Viviescas, M.~Tiersch (Springer, Berlin, 2009), pp. 221--276

\bibitem{Breuer:2002:oq}
H.P. Breuer, F.~Petruccione, \emph{The Theory of Open Quantum Systems} (Oxford
  University Press, Oxford, 2002)

\bibitem{Schlosshauer:2006:rw}
M.~Schlosshauer, A.~Fine, in \emph{Quantum Mechanics at the Crossroads: New
  Perspectives from History, Philosophy and Physics}, ed. by J.~Evans,
  A.~Thorndike (Springer, Berlin, 2006), pp. 125--148

\bibitem{Wooters:1979:az}
W.K. Wootters, W.H. Zurek, Phys. Rev. D \textbf{19}, 473 (1979)

\bibitem{Everett:1957:rw}
H.~Everett, Rev. Mod. Phys. \textbf{29}, 454 (1957)

\bibitem{Einstein:1935:dr}
A.~Einstein, B.~Podolsky, N.~Rosen, Phys. Rev. \textbf{47}, 777 (1935)

\bibitem{Bell:1964:ep}
J.S. Bell, Physics \textbf{1}, 195 (1964)

\bibitem{Bell:1966:ph}
J.S. Bell, Rev. Mod. Phys. \textbf{38}, 447 (1966)

\bibitem{Peres:1978:aa}
A.~Peres, Am. J. Phys. \textbf{46} (1978)

\bibitem{Paz:1993:ta}
J.P. Paz, S.~Habib, W.H. Zurek, Phys. Rev. D \textbf{47}, 488 (1993)

\bibitem{Leggett:1987:pm}
A.J. Leggett, S.~Chakravarty, A.T. Dorsey, M.P.A. Fisher, A.~Garg, Rev. Mod.
  Phys. \textbf{59}, 1 (1987)

\bibitem{Mokarzel:2002:za}
S.G. Mokarzel, A.N. Salgueiro, M.C. Nemes, Phys. Rev. A \textbf{65}, 044101
  (2002)

\bibitem{Hornberger:2003:un}
K.~Hornberger, J.E. Sipe, Phys. Rev. A \textbf{68}, 012105 (2003)

\bibitem{Kraus:1983:ee}
K.~Kraus, \emph{States, Effects, and Operations} (Springer, Berlin, 1983)

\bibitem{Zurek:1991:vv}
W.H. Zurek, Phys. Today \textbf{44}, 36 (1991).
\newblock See also the updated version available as eprint quant-ph/0306072

\bibitem{Gallis:1990:un}
M.R. Gallis, G.N. Fleming, Phys. Rev. A \textbf{42}, 38 (1990)

\bibitem{Diosi:1995:um}
L.~Di{\'o}si, Europhys. Lett. \textbf{30}, 63 (1995)

\bibitem{Hornberger:2006:tb}
K.~Hornberger, Phys. Rev. Lett. \textbf{97}, 060601 (2006)

\bibitem{Hornberger:2008:ii}
K.~Hornberger, B.~Vacchini, Phys. Rev. A \textbf{77}, 022112 (2008)

\bibitem{Busse:2009:aa}
M.~Busse, K.~Hornberger, J. Phys. A: Math. Theor. \textbf{42}, 362001 (2009)

\bibitem{Busse:2010:aa}
M.~Busse, K.~Hornberger, J. Phys. A: Math. Theor. \textbf{43}, 015303 (2010)

\bibitem{Harris:1981:rc}
R.A. Harris, L.~Stodolsky, J. Chem. Phys. \textbf{74}, 2145 (1981)

\bibitem{Zeh:1999:qr}
H.D. Zeh, in \emph{Decoherence: {T}heoretical, Experimental, and Conceptual
  Problems}, ed. by P.~Blanchard, D.~Giulini, E.~Joos, C.~Kiefer,
  I.~Stamatescu, Lecture Notes in Physics {No.\ 538} (Springer, Berlin, 2000),
  pp. 19--42

\bibitem{Paz:1999:vv}
J.P. Paz, W.H. Zurek, Phys. Rev. Lett. \textbf{82}, 5181 (1999)

\bibitem{Zurek:1993:qq}
W.H. Zurek, S.~Habib, J.P. Paz, Phys. Rev. Lett. \textbf{70}, 1187 (1993)

\bibitem{Zurek:1993:pu}
W.H. Zurek, Prog. Theor. Phys. \textbf{89}, 281 (1993)

\bibitem{Hu:1992:om}
B.L. Hu, J.P. Paz, Y.~Zhang, Phys. Rev. D \textbf{45}, 2843 (1992)

\bibitem{Zurek:1998:re}
W.H. Zurek, Philos. Trans. R. Soc. London, Ser. A \textbf{356}, 1793 (1998)

\bibitem{Diosi:2000:yn}
L.~Di{\'o}si, C.~Kiefer, Phys. Rev. Lett. \textbf{85}, 3552 (2000)

\bibitem{Eisert:2003:ib}
J.~Eisert, Phys. Rev. Lett. \textbf{92}, 210401 (2004)

\bibitem{Palma:1996:yy}
G.M. Palma, K.A. Suominen, A.K. Ekert, Proc. R. Soc. Lond. A \textbf{452}, 567
  (1996)

\bibitem{Lidar:1998:uu}
D.A. Lidar, I.L. Chuang, K.B. Whaley, Phys. Rev. Lett. \textbf{81}, 2594 (1998)

\bibitem{Zanardi:1997:yy}
P.~Zanardi, M.~Rasetti, Phys. Rev. Lett. \textbf{79}, 3306 (1997)

\bibitem{Zanardi:1997:tv}
P.~Zanardi, M.~Rasetti, Mod. Phys. Lett. B \textbf{11}, 1085 (1997)

\bibitem{Zanardi:1998:oo}
P.~Zanardi, Phys. Rev. A \textbf{57}, 3276 (1998)

\bibitem{Lidar:1999:fa}
D.A. Lidar, D.~Bacon, K.B. Whaley, Phys. Rev. Lett. \textbf{82}, 4556 (1999)

\bibitem{Bacon:2000:yy}
D.~Bacon, J.~Kempe, D.A. Lidar, K.B. Whaley, Phys. Rev. Lett. \textbf{85}, 1758
  (2000)

\bibitem{Duan:1998:yb}
L.M. Duan, G.C. Guo, Phys. Rev. A \textbf{57}, 737 (1998)

\bibitem{Zanardi:2001:oo}
P.~Zanardi, Phys. Rev. A \textbf{63}, 012301 (2001)

\bibitem{Knill:2000:aa}
E.~Knill, R.~Laflamme, L.~Viola, Phys. Rev. Lett. \textbf{82}, 2525 (2000)

\bibitem{Kempe:2001:oo}
J.~Kempe, D.~Bacon, D.A. Lidar, K.B. Whaley, Phys. Rev. A \textbf{63}, 042307
  (2001)

\bibitem{Lidar:2003:aa}
D.A. Lidar, K.B. Whaley, in \emph{Irreversible Quantum Dynamics},
  \emph{Springer Lecture Notes in Physics}, vol. 622, ed. by F.~Benatti,
  R.~Floreanini (Springer, Berlin, 2003), pp. 83--120.
\newblock Also available as eprint quant-ph/0301032

\bibitem{Zurek:2003:pl}
W.H. Zurek, in \emph{Science and Ultimate Reality}, ed. by J.D. Barrow, P.C.W.
  Davies, C.H. Harper (Cambridge University Press, Cambridge, England, 2004),
  pp. 121--137

\bibitem{Ollivier:2003:za}
H.~Ollivier, D.~Poulin, W.H. Zurek, Phys. Rev. Lett. \textbf{93}, 220401 (2004)

\bibitem{Ollivier:2004:im}
H.~Ollivier, D.~Poulin, W.H. Zurek, Phys. Rev. A \textbf{72}, 042113 (2005)

\bibitem{Blume:2004:oo}
R.~Blume-Kohout, W.H. Zurek, Found. Phys. \textbf{35}, 1857 (2005)

\bibitem{Blume:2005:oo}
R.~Blume-Kohout, W.H. Zurek, Phys. Rev. A \textbf{73}, 062310 (2006)

\bibitem{Zurek:2009:om}
W.H. Zurek, Nature Phys. \textbf{5}, 181 (2009)

\bibitem{Riedel:2010:un}
C.J. Riedel, W.H. Zurek, Phys. Rev. Lett. \textbf{105}, 020404 (2010)

\bibitem{Riedel:2011:un}
C.J. Riedel, W.H. Zurek, New J. Phys. \textbf{13}, 073038 (2011)

\bibitem{Riedel:2012:un}
C.J. Riedel, W.H. Zurek, M.~Zwolak, New J. Phys. \textbf{14}, 083010 (2012)

\bibitem{Zwolak:2014:tt}
M.~Zwolak, C.J. Riedel, W.H. Zurek, Phys. Rev. Lett. \textbf{112}, 140406
  (2014)

\bibitem{Blume:2007:oo}
R.~Blume-Kohout, W.H. Zurek, Phys. Rev. Lett. \textbf{101}, 240405 (2008)

\bibitem{Ollivier:2001:az}
H.~Ollivier, W.H. Zurek, Phys. Rev. Lett. \textbf{88}, 017901 (2002)

\bibitem{Schneider:1998:yz}
S.~Schneider, G.J. Milburn, Phys. Rev. A \textbf{57}, 3748 (1998)

\bibitem{Miquel:1997:zz}
C.~Miquel, J.P. Paz, W.H. Zurek, Phys. Rev. Lett. \textbf{78}, 3971 (1997)

\bibitem{Vandersypen:2004:ra}
L.M.K. Vandersypen, I.L. Chuang, Rev. Mod. Phys. \textbf{76}, 1037 (2004)

\bibitem{Martinis:2003:bz}
J.M. Martinis, S.~Nam, J.~Aumentado, K.M. Lang, C.~Urbina, Phys. Rev. B
  \textbf{67}, 094510 (2003)

\bibitem{Myatt:2000:yy}
C.J. Myatt, B.E. King, Q.A. Turchette, C.A. Sackett, D.~Kielpinski, W.M. Itano,
  C.~Monroe, D.J. Wineland, Nature \textbf{403}, 269 (2000)

\bibitem{Schneider:1999:tt}
S.~Schneider, G.J. Milburn, Phys. Rev. A \textbf{59}, 3766 (1999)

\bibitem{Turchette:2000:aa}
Q.A. Turchette, C.J. Myatt, B.E. King, C.A. Sackett, D.~Kielpinski, W.M. Itano,
  C.~Monroe, D.J. Wineland, Phys. Rev. A \textbf{62}, 053807 (2000)

\bibitem{Kraus:1971:ii}
K.~Kraus, Ann. Phys. \textbf{64}, 311 (1971)

\bibitem{Gorini:1976:tt}
V.~Gorini, A.~Kossakowski, E.C.G. Sudarshan, J. Math. Phys. \textbf{17}, 821
  (1976)

\bibitem{Lindblad:1976:um}
G.~Lindblad, Commun. Math. Phys. \textbf{48}, 119 (1976)

\bibitem{Alicki:2001:aa}
R.~Alicki, M.~Fannes, \emph{Quantum Dynamical Systems} (Oxford University
  Press, Oxford, 2001)

\bibitem{Alicki:2007:uu}
R.~Alicki, K.~Lendi, \emph{Quantum Dynamical Semigroups and Applications},
  \emph{Lect. Notes Phys.}, vol. 717, 2nd edn. (Springer, Berlin/Heidelberg,
  2007)

\bibitem{Benatti:2005:ii}
F.~Benatti, R.~Floreanini, Int. J. Mod. Phys. B \textbf{19}, 3063 (2005).
\newblock \doi{10.1142/S0217979205032097}

\bibitem{Dumke:1979:ia}
R.~D{\"u}mcke, H.~Spohn, Z. Phys. B \textbf{34}, 419 (1979)

\bibitem{Gorini:1978:uf}
V.~Gorini, A.~Frigerio, M.~Verri, A.~Kossakowski, E.C.G. Sudarshan, Rep. Math.
  Phys. \textbf{13}, 149 (1978)

\bibitem{Davies:1974:tw}
E.B. Davies, Commun. Math. Phys. \textbf{39}, 91 (1974)

\bibitem{Kossakowski:1972:tf}
A.~Kossakowski, Rep. Math. Phys. \textbf{3}, 247 (1972)

\bibitem{Barchielli:1991:fv}
A.~Barchielli, V.P. Belavkin, J. Phys. A: Math. Gen. \textbf{24}, 1495 (1991)

\bibitem{Belavkin:1989:an}
V.P. Belavkin, in \emph{Lecture Notes in Control and Information Sciences},
  vol. 121 (Springer, Berlin, 1989), pp. 245--265

\bibitem{Belavkin:1989:am}
V.P. Belavkin, J. Phys. A: Math. Gen. \textbf{22}, L1109 (1989)

\bibitem{Belavkin:1989:um}
V.P. Belavkin, Phys. Lett. A \textbf{140}, 355 (1989)

\bibitem{Belavkin:1995:tt}
V.P. Belavkin, in \emph{Chaos: The Interplay Between Stochastic and
  Deterministic Behaviour}, ed. by P.~Garbaczewksi, M.~Wolf, A.~Veron, Lecture
  Notes in Physics (Springer, 1995), pp. 21--41

\bibitem{Diosi:1988:wx}
L.~Di{\'o}si, Phys. Lett. A \textbf{129}, 419 (1988)

\bibitem{Diosi:1988:hn}
L.~Di{\'o}si, Phys. Lett. A \textbf{132}, 233 (1988)

\bibitem{Diosi:1988:bv}
L.~Di{\'o}si, J. Phys. A \textbf{21}, 2885 (1988)

\bibitem{Gisin:1984:qs}
N.~Gisin, Phys. Rev. Lett. \textbf{52}, 1657 (1984)

\bibitem{Gisin:1989:jn}
N.~Gisin, Helv. Phys. Acta \textbf{62}, 363 (1989)

\bibitem{Wiseman:1994:qq}
H.M. Wiseman, Phys. Rev. A \textbf{49}, 2133 (1994)

\bibitem{Goan:2001:rz}
H.S. Goan, G.J. Milburn, H.M. Wiseman, H.B. Sun, Phys. Rev. B \textbf{63},
  125326 (2001)

\bibitem{Plenio:1998:bb}
M.B. Plenio, P.L. Knight, Rev. Mod. Phys. \textbf{70}, 101 (1998)

\bibitem{Prokofev:2000:zz}
N.V. Prokof'ev, P.C.E. Stamp, Rep. Prog. Phys. \textbf{63}, 669 (2000)

\bibitem{Dube:2001:zz}
M.~Dub{\'e}, P.C.E. Stamp, Chem. Phys. \textbf{268}, 257 (2001)

\bibitem{Groeblacher:2013:im}
S.~Gr{\"o}blacher, A.~Trubarov, N.~Prigge, M.~Aspelmeyer, J.~Eisert, Nature
  Comm. \textbf{6}, 7606 (2015)

\bibitem{Chaturvedi:1979:pm}
S.~Chaturvedi, F.~Shibata, Z. Phys. B \textbf{35}, 297 (1979)

\bibitem{Shibata:1980:ma}
F.~Shibata, T.~Arimitsu, J. Phys. Soc. Jpn. \textbf{49}, 891 (1980)

\bibitem{Royer:1972:um}
A.~Royer, Phys. Rev. A \textbf{6}, 1741 (1972)

\bibitem{Royer:2003:za}
A.~Royer, Phys. Lett. A \textbf{315}, 335 (2003)

\bibitem{Feynman:1963:jj}
R.~Feynman, F.L. Vernon, Ann. Phys. (N.Y.) \textbf{24}, 118 (1963)

\bibitem{Caldeira:1983:gv}
A.~Caldeira, A.~Leggett, Ann. Phys. (N.Y.) \textbf{149}, 374 (1983)

\bibitem{Lounasmaa:1974:yb}
O.V. Lounasmaa, \emph{Experimental Principles and Methods below 1 K} (Academic
  Press, New York, 1974)

\bibitem{Adler:2006:yb}
S.L. Adler, J. Phys. A: Math. Gen. \textbf{39}, 14067 (2006)

\bibitem{Tegmark:1993:uz}
M.~Tegmark, Found. Phys. Lett. \textbf{6}, 571 (1993)

\bibitem{Hackermuller:2003:uu}
L.~Hackerm{\"u}ller, K.~Hornberger, B.~Brezger, A.~Zeilinger, M.~Arndt, Appl.
  Phys. B \textbf{77}, 781 (2003)

\bibitem{Hornberger:2003:tv}
K.~Hornberger, S.~Uttenthaler, B.~Brezger, L.~Hackerm{\"u}ller, M.~Arndt,
  A.~Zeilinger, Phys. Rev. Lett. \textbf{90}, 160401 (2003)

\bibitem{Hornberger:2004:bb}
K.~Hornberger, J.E. Sipe, M.~Arndt, Phys. Rev. A \textbf{70}, 053608 (2004)

\bibitem{Nimmrichter:2011:pr}
S.~Nimmrichter, K.~Hornberger, P.~Haslinger, M.~Arndt, Phys. Rev. A
  \textbf{83}, 043621 (2011)

\bibitem{Kokorowski:2001:ub}
D.A. Kokorowski, A.D. Cronin, T.D. Roberts, D.E. Pritchard, Phys. Rev. Lett.
  \textbf{86}, 2191 (2001)

\bibitem{Uys:2005:yb}
H.~Uys, J.D. Perreault, A.D. Cronin, Phys. Rev. Lett. \textbf{95}, 150403
  (2005)

\bibitem{Caldeira:1983:on}
A.O. Caldeira, A.J. Leggett, Physica A \textbf{121}, 587 (1983)

\bibitem{Walls:1985:lm}
D.F. Walls, M.J. Collett, G.J. Milburn, Phys. Rev. D \textbf{32}, 3208 (1985)

\bibitem{Weiss:1999:tv}
U.~Weiss, \emph{Quantum Dissipative Systems} (World Scientific, Singapore,
  1999)

\bibitem{Schlosshauer:2008:os}
M.~Schlosshauer, A.P. Hines, G.J. Milburn, Phys. Rev. A \textbf{77}, 022111
  (2008)

\bibitem{Stamp:1998:im}
P.C.E. Stamp, in \emph{Tunnelling in Complex Systems}, ed. by S.~Tomsovic
  (World Scientific, Singapore, 1998), pp. 101--197

\bibitem{Prokofev:1995:ab}
N.V. Prokof{'}ev, P.C.E. Stamp,   (1995)

\bibitem{Prokofev:1993:aa}
N.V. Prokof{'}ev, P.C.E. Stamp, J. Phys. Chem. Lett. \textbf{5}, L663 (1993)

\bibitem{Dobrovitski:2003:az}
V.V. Dobrovitski, H.A. {De Raedt}, M.I. Katsnelson, B.N. Harmon, Phys. Rev.
  Lett. \textbf{90}, 210401 (2003).
\newblock \doi{10.1103/PhysRevLett.90.210401}

\bibitem{Cucchietti:2005:om}
F.M. Cucchietti, J.P. Paz, W.H. Zurek, Phys. Rev. A \textbf{72}, 052113 (2005).
\newblock \doi{10.1103/PhysRevA.72.052113}

\bibitem{Caldeira:1993:bz}
A.O. Caldeira, A.H. {Castro Neto}, T.O. de~Carvalho, Phys. Rev. B \textbf{48},
  13974 (1993)

\bibitem{Miquel:1996:ra}
C.~Miquel, J.P. Paz, R.~Perazzo, Phys. Rev. A \textbf{54}, 2605 (1996)

\bibitem{Steane:1996:cd}
A.M. Steane, Phys. Rev. Lett. \textbf{77}, 793 (1996)

\bibitem{Shor:1995:rx}
P.W. Shor, Phys. Rev. A \textbf{52}, R2493 (1995)

\bibitem{Steane:2001:dx}
A.M. Steane, in \emph{Decoherence and Its Implications in Quantum Computation
  and Information Transfer}, ed. by P.~Turchi, A.~Gonis (IOS Press, Amsterdam,
  2001), pp. 284--298.
\newblock Also available as eprint quant-ph/0304016

\bibitem{Knill:2002:rx}
E.~Knill, R.~Laflamme, A.~Ashikhmin, H.~Barnum, L.~Viola, W.~Zurek, LA Science
  \textbf{27}, 188 (2002)

\bibitem{Nielsen:2000:tt}
M.A. Nielsen, I.L. Chuang, \emph{Quantum Computation and Quantum Information}
  (Cambridge University Press, Cambridge, 2000)

\bibitem{Reina:2002:ta}
J.H. Reina, L.~Quiroga, N.F. Johnson, Phys. Rev. A 65 \textbf{65}, 032326
  (2002)

\bibitem{Bacon:1999:aq}
D.~Bacon, D.A. Lidar, K.B. Whaley, Phys. Rev. A \textbf{60}, 1944 (1999)

\bibitem{Lidar:2001:oo}
D.A. Lidar, D.~Bacon, J.~Kempe, K.B. Whaley, Phys. Rev. A \textbf{63}, 022307
  (2001)

\bibitem{Dalvit:2000:bb}
D.A.R. Dalvit, J.~Dziarmaga, W.H. Zurek, Phys. Rev. A \textbf{62}, 013607
  (2000)

\bibitem{Poyatos:1996:um}
J.F. Poyatos, J.I. Cirac, P.~Zoller, Phys. Rev. Lett. \textbf{77}, 4728 (1996)

\bibitem{Carvalho:2001:ua}
A.R.R. Carvalho, P.~Milman, R.L. de~Matos~Filho, L.~Davidovich, Phys. Rev.
  Lett. \textbf{86}, 4988 (2001)

\bibitem{Viola:1998:uu}
L.~Viola, S.~Lloyd, Phys. Rev. A \textbf{58}, 2733 (1998)

\bibitem{Viola:1999:zp}
L.~Viola, E.~Knill, S.~Lloyd, Phys. Rev. Lett. \textbf{82}, 2417 (1999)

\bibitem{Zanardi:1999:oo}
P.~Zanardi, Phys. Lett. A \textbf{258}, 77 (1999)

\bibitem{Viola:2000:pp}
L.~Viola, E.~Knill, S.~Lloyd, Phys. Rev. Lett. \textbf{85}, 3520 (2000)

\bibitem{Wu:2002:aa}
L.A. Wu, D.A. Lidar, Phys. Rev. Lett. \textbf{88}, 207902 (2002)

\bibitem{Wu:2002:bb}
L.A. Wu, M.S. Byrd, D.A. Lidar, Phys. Rev. Lett. \textbf{89}, 127901 (2002)

\bibitem{Arndt:2005:mi}
M.~Arndt, K.~Hornberger, A.~Zeilinger, Phys. World \textbf{18}, 35 (2005)

\bibitem{Kaiser:2001:tm}
R.~Kaiser, C.~Westbrook, F.~David (eds.).
\newblock \emph{Coherent Atomic Matter Waves, Les Houches Session LXXII}, Les
  Houches Summer School Series (Springer, Berlin, 2001)

\bibitem{Blencowe:2004:mm}
M.~Blencowe, Phys. Rep. \textbf{395}, 159 (2004)

\bibitem{Aspelmeyer:2013:aa}
M.~Aspelmeyer, T.J. Kippenberg, F.~Marquardt, Rev. Mod. Phys. \textbf{86}, 1391
  (2014)

\bibitem{Bassi:2003:yb}
A.~Bassi, G.C. Ghirardi, Phys. Rep. \textbf{379}, 257 (2003)

\bibitem{Adler:2007:um}
S.L. Adler, J. Phys. A \textbf{40}, 2935 (2007)

\bibitem{Bassi:2010:aa}
A.~Bassi, D.A. Deckert, L.~Ferialdi, EPL \textbf{92}, 50006 (2010)

\bibitem{Nimmrichter:2013:aa}
S.~Nimmrichter, K.~Hornberger, Phys. Rev. Lett. \textbf{110}, 160403 (2013)

\bibitem{Brune:1996:om}
M.~Brune, E.~Hagley, J.~Dreyer, X.~Ma{\^i}tre, A.~Maali, C.~Wunderlich, J.M.
  Raimond, S.~Haroche, Phys. Rev. Lett. \textbf{77}, 4887 (1996)

\bibitem{Davidovich:1996:sa}
L.~Davidovich, M.~Brune, J.M. Raimond, S.~Haroche, Phys. Rev. A \textbf{53},
  1295 (1996)

\bibitem{Maitre:1997:tv}
X.~Ma{\^i}tre, E.~Hagley, J.~Dreyer, A.~Maali, C.W.M. Brune, J.M. Raimond,
  S.~Haroche, J. Mod. Opt. \textbf{44}, 2023 (1997)

\bibitem{Auffeves:2003:za}
A.~Auffeves, P.~Maioli, T.~Meunier, S.~Gleyzes, G.~Nogues, M.~Brune, J.M.
  Raimond, S.~Haroche, Phys. Rev. Lett. \textbf{91}, 230405 (2003)

\bibitem{Deleglise:2008:oo}
S.~Del{\'e}glise, I.~Dotsenko, C.~Sayrin, J.~Bernu, M.~Brune, J.M. Raimond,
  S.~Haroche, Nature \textbf{455}, 510 (2008)

\bibitem{Arndt:1999:rc}
M.~Arndt, O.~Nairz, J.~Vos-Andreae, C.~Keller, G.~van~der Zouw, A.~Zeilinger,
  Nature \textbf{401}, 680 (1999)

\bibitem{Gerlich:2011:aa}
S.~Gerlich, S.~Eibenberger, M.~Tomandl, S.~Nimmrichter, K.~Hornberger, P.J.
  Fagan, J.~T{\"u}xen, M.~Mayor, M.~Arndt, Nature Comm. \textbf{2}, 263 (2012)

\bibitem{Eibenberger:2013:az}
S.~Eibenberger, S.~Gerlich, M.~Arndt, M.~Mayor, J.~T{\"u}xen, Phys. Chem. Chem.
  Phys. \textbf{15}, 14696 (2013)

\bibitem{Gerlich:2007:om}
S.~Gerlich, L.~Hackerm{\"u}ller, K.~Hornberger, A.~Stibor, H.~Ulbricht,
  F.~Goldfarb, T.~Savas, M.~M{\"u}ri, M.~Mayor, M.~Arndt, Nature Phys.
  \textbf{3}, 711 (2007)

\bibitem{Haslinger:2013:ii}
P.~Haslinger, N.~D{\"o}rre, P.~Geyer, J.~Rodewald, S.~Nimmrichter, M.~Arndt,
  Nature Phys. \textbf{9}, 144 (2013)

\bibitem{Hornberger:2005:mo}
K.~Hornberger, L.~Hackerm{\"u}ller, M.~Arndt, Phys. Rev. A \textbf{71}, 023601
  (2005)

\bibitem{Hackermuller:2004:rd}
L.~Hackerm{\"u}ller, K.~Hornberger, B.~Brezger, A.~Zeilinger, M.~Arndt, Nature
  \textbf{427}, 711 (2004)

\bibitem{Arndt:2002:bo}
M.~Arndt, O.~Nairz, A.~Zeilinger, in \emph{Quantum [Un]Speakables: From Bell to
  Quantum Information}, ed. by R.A. Bertlmann, A.~Zeilinger (Springer, Berlin,
  2002), pp. 333--351

\bibitem{Hornberger:2006:tx}
K.~Hornberger, Phys. Rev. A \textbf{73}, 052102 (2006)

\bibitem{Friedman:2000:rr}
J.R. Friedman, V.~Patel, W.~Chen, S.K. Yolpygo, J.E. Lukens, Nature
  \textbf{406}, 43 (2000)

\bibitem{Wal:2000:om}
C.H. van~der Wal, A.C.J. ter Haar, F.K. Wilhelm, R.N. Schouten, C.J.P.M.
  Harmans, T.P. Orlando, S.~Lloyd, J.E. Mooij, Science \textbf{290}, 773 (2000)

\bibitem{Chiorescu:2003:ta}
I.~Chiorescu, Y.~Nakamura, C.J.P.M. Harmans, J.E. Mooij, Science \textbf{21},
  1869 (2003)

\bibitem{Bertet:2005:un}
P.~Bertet, I.~Chiorescu, G.~Burkard, K.~Semba, C.J.P.M. Harmans, D.P.
  DiVincenzo, J.E. Mooij, Phys. Rev. Lett. \textbf{95}, 257002 (2005)

\bibitem{Nakamura:1999:ub}
Y.~Nakamura, Y.A. Pashkin, J.S. Tsai, Nature \textbf{398}, 786 (1999)

\bibitem{Vion:2002:oo}
D.~Vion, A.~Aassime, A.~Cottet, P.~Joyez, H.~Pothier, C.~Urbina, D.~Esteve,
  M.H. Devoret, Science \textbf{296}, 886 (2002)

\bibitem{Yu:2002:yb}
Y.~Yu, S.~Han, X.~Chu, S.I. Chu, Z.~Wang, Science \textbf{296}, 889 (2002)

\bibitem{Martinis:2002:qq}
J.M. Martinis, S.~Nam, J.~Aumentado, C.~Urbina, Phys. Rev. Lett. \textbf{89},
  117901 (2002)

\bibitem{Camilleri:2009:aq}
K.~Camilleri, Stud. Hist. Phil. Mod. Phys. \textbf{40}, 290 (2009)

\bibitem{Wallace:2010:im}
D.~Wallace, in \emph{Many Worlds? {E}verett, Quantum Theory and Reality}, ed.
  by S.~Saunders, J.~Barrett, A.~Kent, D.~Wallace (Oxford University Press,
  Oxford, 2010), pp. 53--72

\end{thebibliography}

\end{document}